\documentclass[twocolumn]{aastex631}

\pdfoutput=1 
\usepackage{amsmath,amstext}
\usepackage[T1]{fontenc}
\usepackage{gensymb}
\usepackage{subfigure}
\usepackage{longtable}

\usepackage{subfigure}
\usepackage{savesym}
\savesymbol{tablenum}
\usepackage{siunitx}
\restoresymbol{SIX}{tablenum}
\DeclareSIUnit{\year}{yr}
\DeclareSIUnit{\mas}{mas}
\DeclareSIUnit{\astronomicalunit}{au}
\DeclareSIUnit{\parsec}{pc}

\shorttitle{Unresolved Companions in the Lobster Diagram}
\shortauthors{Hartman et al.}
\graphicspath{{./}{figures/}}

\begin{document}

\title{Vetting the Lobster Diagram: Searching for Unseen Companions in Wide Binaries using NASA Space Exoplanet Missions}

\author[0000-0003-4236-6927]{Zachary D. Hartman}
\affiliation{Gemini Observatory/NSF's NOIRLab, 670 A'ohoku Place, Hilo, HI 96720, USA}

\author[0000-0002-2437-2947]{S\'{e}bastien L\'{e}pine}
\affiliation{Georgia State University, Department of Physics \& Astronomy, Georgia State University, 25 Park Place South, Suite 605, Atlanta GA, 30303, USA}

\author[0000-0003-3410-5794]{Ilija Medan}
\affiliation{Georgia State University, Department of Physics \& Astronomy, Georgia State University, 25 Park Place South, Suite 605, Atlanta GA, 30303, USA}


\begin{abstract}
Over the past decade, the number of known wide binary systems has exponentially expanded thanks to the release of data from the \textit{Gaia} Mission.  Some of these wide binary systems are actually higher-order multiples, where one of the components is an unresolved binary itself.  One way to search for these systems is by identifying overluminous components in the systems.  In this study, we examine 4947 K+K wide binary pairs from the SUPERWIDE catalog and quantify the relative color and luminosity of the components to find evidence for additional, unresolved companions.  The method is best illustrated in a graph we call the ``Lobster'' diagram.  To confirm that the identified overluminous components are close binary systems, we cross-match our wide binaries with the TESS, K2 and Kepler archives and search for the signs of eclipses and fast stellar rotation modulation in the light curves.  We find that $78.9\%\pm20.7\%$ of the wide binaries which contain an eclipsing system are identified to be overluminous in the ``Lobster Diagram'' and $73.5\%\pm12.4\%$ of the wide binaries which contain a component showing fast rotation ($P<5$) days also show an overluminous component.  From these results, we calculate a revised lower limit on the higher-order multiplicity fraction for K+K wide binaries of $40.0\%\pm1.6\%$.  We also examine the higher-order multiplicity fraction as a function of projected physical separation and metallicity.  The fraction is unusually constant as a function of projected physical separation while we see no statistically significant evidence that the fraction varies with metallicity.
\end{abstract}

\section{Introduction}
Binary stellar systems appear in a wide variety of forms and configurations.  Wide binary systems can have separations that can reach many thousands of au and are found mostly as visual binaries where both components are easily resolved on the sky.  Angular separations, proper motion differences, parallax/distance differences and radial velocity differences have been used in various different combinations to find and confirm these systems in large astrometric stellar catalogs \citep{2004chanamedarkmatter,2007lepinewide,2010dhitalslowpokes1,2011lepine,2011shaya,2012tokovinin,2014tokovininfgsample1,2015dhitalslowpokes2,2016deacon,2017andrews,2017oh,2017oelkers,2018elbadry,2018coronado,2019jimenez,2020hartman,2021elbadry}.  On the other hand, unresolved, close binary systems with separations reaching to less than 1 au can be more challenging to identify and are found using a wider variety of discovery methods, including using high resolution imaging, spectroscopy and photometry. 

It has been well established for solar-type stars that around half of the wide binaries are higher-order multiples, i.e. triples, quadruples, etc. \citep{2010raghavan,2014tokovininfgsample2,2019moemetal}.  However, the pathway by which the widest of these systems originated has not been well established.  Three scenarios for how these systems form include the unfolding of triple systems \citep{2012reipurth}, the binding of stars during the cluster dissolution phase \citep{2010moeckel,2010kouwenhoven}, and the pairings of adjacent cores in star forming regions \citep{2017corestokovinin}.  These scenarios address wide binary formation at the largest separations ($>10,000$ au).  At shorter separations, much work has already gone into examining close binary formation by characterizing their statistical properties such as the binary fraction as a function of metallicity \citep{2019moemetal,2019elbadry,2021hwang}.  

One feature that each of the three scenarios predicts is that wide binaries should initially have a large fraction of higher-order multiples, and this fraction is expected to increase over time due to the ability that multiple systems have to better survive interactions with the local Galactic environment.  Data from \cite{2010raghavan} indeed shows that for the eleven systems that have separations larger than 10,000 au, ten are higher-order multiples.  Unfortunately, most studies that have measured higher-order multiplicity fractions are for solar-type (F-, G-, and early K-dwarf) systems.  \citet{2010law} examined the higher order multiplicity for M+M wide binaries and also found that the higher order multiplicity fraction increases as a function of projected physical separation; however, this was only on a sample of 36 wide binaries and extended only out to orbital separations of 6500 au.  Knowing the value of this higher order multiplicity fraction over a broader range of spectral types and orbital separations will help determine if there is a formation scenario that is dominant or if wide binary formation is a mixture of different processes.  

Searching a large number of wide binary systems for additional close companions is a challenging task, especially if the companions are unresolved.  For \textit{Gaia}, \citet{2018roboaogaiacheck} found that the resolution limit was between 0.7\arcsec and 1\arcsec.  For most systems with separations below this, resource-intensive follow-up, such as high resolution imaging or spectroscopy, are normally needed to find companions.  However, one particular discovery method for finding unresolved close binary systems that does not require a large commitment of resources is looking for overluminous stars in the \textit{Gaia} catalog itself.  There are, however, several reasons that may cause a star to appear overluminous compared to a main-sequence star of similar color.  The star could be young and still contracting on its way to the main sequence phase, it could be a highly active star in a flaring state, or it could be an unresolved binary.  In most of the H-R diagram and even along the main-sequence, it is generally difficult to identify overluminous systems due to what is called the ``cosmic scatter'', i.e. to variations in the luminosity and color of main-sequence stars due to differences in age, metallicity, and state of activity.  However, an examination of the \textit{Gaia} H-R diagram shows a very suggestive doubling of the main sequence in the K- and M- dwarf regime that is a clear indication of a significant population of unresolved binaries - and provides the potential means to identify them.  

In our first paper \citep{2020hartman}, we examined the scatter in the color-magnitude relationship for K-dwarfs in the \textit{Gaia} H-R diagram  using a subset of 2227 K+K wide binary systems.  We first defined a fiducial line running parallel to the mean color-magnitude trend, and calculated an ``overluminosity factor,'' which is defined as the vertical magnitude offset from the fiducial line for any star.  For a majority of the stars in common proper motion pairs, we found a strong correlation between the overluminosity of the primary and that of the secondary, consistent with the idea that much of the ``cosmic scatter'' is due to metallicity differences in K-dwarfs, and that K+K systems are chemically homogeneous, with both components expected to show the same offset from the fiducial main-sequence.  Exploiting this, we devised a simple but useful tool for finding potential unresolved close binaries that are part of wide systems: these unresolved pairs are identified if either one of the wide components appears to be significantly overluminous compared to the other.  Unresolved pairs are most easily identified as outliers in the so-called ``Lobster Diagram,'' which plots the overluminosity factor of the primary as a function of the secondary's factor.

In this paper, we widen our sample to 4947 wide binaries where both components have \textit{Gaia} $G_{BP} - G_{RP}$ colors consistent with being K-dwarfs, all pairs have a probability of being gravitationally bound systems $>95\%$, and the distance limit we had in place in \cite{2020hartman} was removed.  To confirm that the ``Lobster diagram'' is indeed identifying unresolved close companions, we further examine light curves from the TESS, K2 and Kepler surveys searching for signs of light curve modulations indicative of close/eclipsing binaries.  We find convincing evidence that the method accurately identifies unresolved components, provided that their binary mass ratio q= $M_{sec}/M_{pri}$ is larger than 0.5.  Finally, we estimate the higher-order multiplicity fraction for the whole sample and search for variations of this fraction as a function of projected physical separation and metallicity.

\section{Data Retrieval and Light Curve Analysis}

\subsection{Wide Binary Identification}
\begin{figure}
\centering
\subfigure{
\includegraphics[width=3.3in]{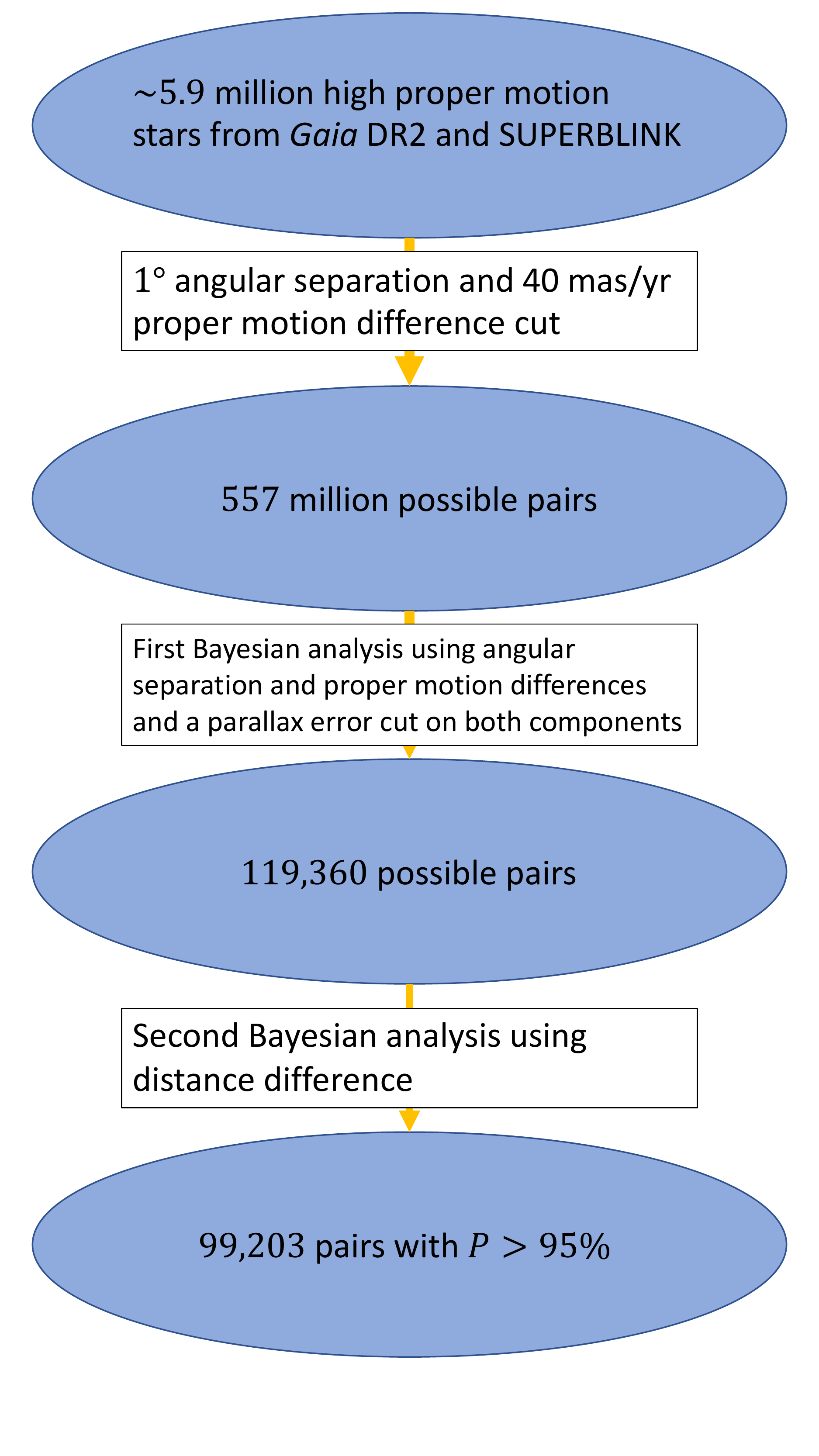}}
\caption{\label{fig:method}Flow chart demonstrating the method used to construct the SUPERWIDE catalog.  Starting with a sample of $\sim5.9$ million high proper motion stars, we conduct a two-step Bayesian analysis to determine which stars are part of wide binaries.  This analysis takes into account angular separations, proper motion differences and distance differences.}
\end{figure}

The method used to identify our sample of wide binaries is detailed in \citet{2020hartman}.  We provide a brief summary here.  Starting with the complete set of $\sim5.9$ million high proper motion stars ($>$\SI{40}{\mas\per\year}) in \textit{Gaia} Data Release 2 \citep{2018gaiadr2,2018lindegrengaia} supplemented by a modest number of additional high proper motion stars from the SUPERBLINK catalog, but not listed in DR2, we conduct a two stage Bayesian analysis that calculates the probability of any two stars to be physical pairs (as opposed to chance alignments) based on their angular separations, proper motion differences and distance/parallax differences.  We present a flow chart of the process in Figure \ref{fig:method}.  The first stage takes the angular separations and proper motion differences and uses empirical model distributions of these two parameters for both chance alignments and real binaries to calculate a first pass real-binary probability.  We then keep only pairs with a first-pass probability greater than $10$\% and with a parallax error on both components less than $10\%$, and run another Bayesian analysis using just the difference in the distances between the components in the pairs, again using empirical model distributions for real binaries and chance alignments.  The end result is the SUPERWIDE catalog listing 99,203 high proper motion pairs with probabilities of being gravitationally bound $>95\%$.

\begin{figure}
\centering
\subfigure{
\includegraphics[width=3.3in]{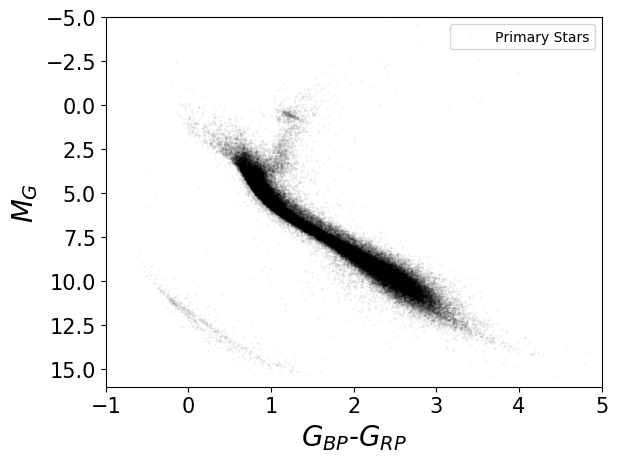}}
\subfigure{
\includegraphics[width=3.3in]{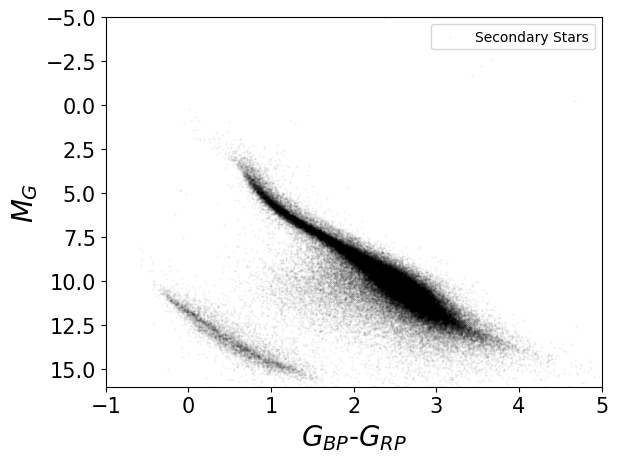}}
\caption{\label{fig:priandsechr}Color-magnitude diagrams for the 99,203 pairs with Bayesian probabilities $>95\%$ of being wide physical binaries.  Top: Color-magnitude diagram for the primary components.  Bottom: Color-magnitude diagram for the secondary components. Primary stars are found of all types, including notable subsets of red giants, subgiants, more massive main-sequence stars, and white dwarfs. Secondaries are overwhelmingly low-mass stars and white dwarfs.  One notable feature is the vertical ``thickening'' of the main sequence for low-mass M stars (\textit{Gaia} $G_{BP} - G_{RP}$  $> 2$).}
\end{figure}

The color-magnitude diagrams for these high probability pairs are presented in Figure \ref{fig:priandsechr}.  Primary stars are shown in the top panel while secondaries are shown on the bottom panel, with ``primary'' and ``secondary'' being determined by brighter $G$ magnitude.  For both panels, the main sequence is well defined and in the color range of the K-dwarfs, $G_{BP} - G_{RP}$ from 1.01 to 1.81, a doubling of the main sequence can be distinguished as a secondary sequence of objects vertically shifted up by $\sim0.7$ magnitudes, looking much like a ``halo'' above the standard main sequence.  This doubling is believed to be caused by the presence of unresolved companions.  While young stars can cause a similar effect as they fall onto the main sequence, as they have yet to fully contract and remain overluminous due to their larger sizes compared to normal main sequence stars, we note that the SUPERWIDE catalog was constructed from a high proper motion sample which is biased against young, field stars, which typically have low relative motions to the Sun and thus are under-represented in high proper motion subsets.  

\begin{figure}
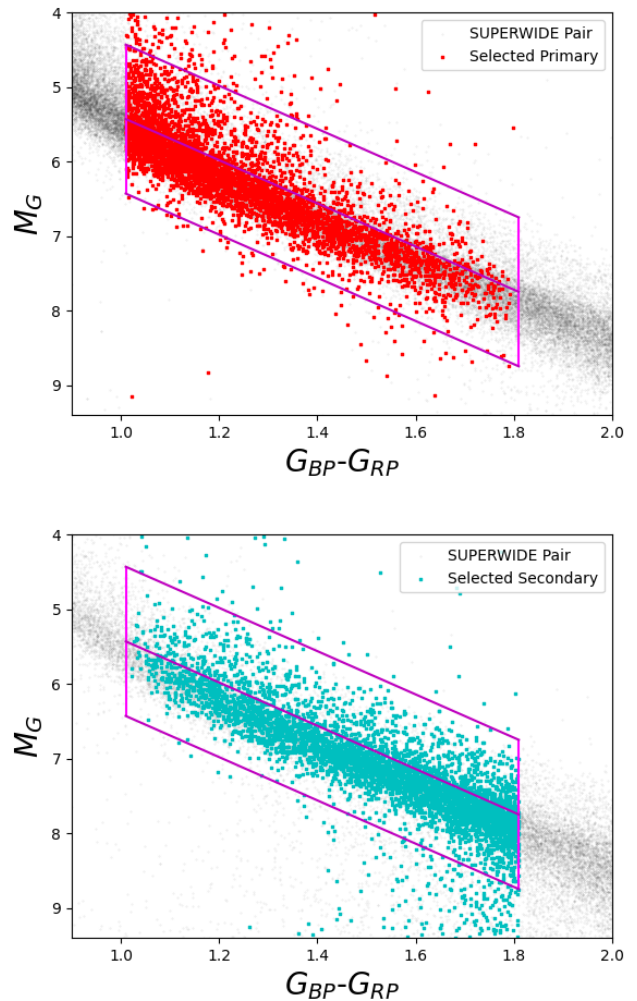

\centering
\subfigure{
\includegraphics[width=3.3in]{../figures/PriHR_zoomedk.png}}
\subfigure{
\includegraphics[width=3.3in]{../figures/SecHR_zoomedk.png}}
\caption{\label{fig:priandsecselection}H-R Diagram for SUPERWIDE with the K+K wide binary selection shown.  The magenta box defines the K dwarf main sequence plus overluminous stars.  The line going through the middle represents the arbitrary reference line $[M_G]_{Kref}$ that is used to define the overluminosity factor for each wide binary.  Red points show the primary stars and the cyan points show the secondary stars selected for our light curve search.  }
\end{figure}

\begin{figure}
\centering
\includegraphics[width=3.3in]{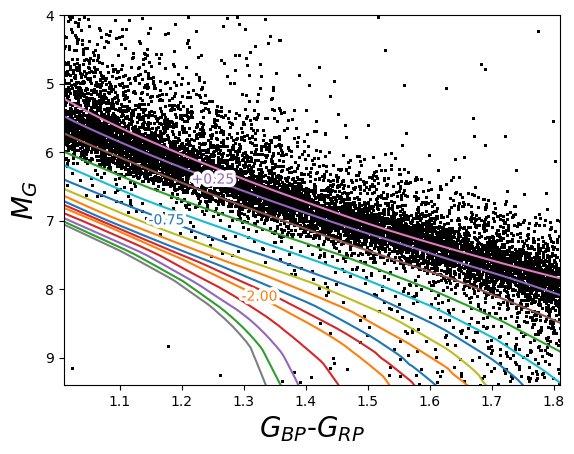}
\caption{\label{fig:overlumhr}Color-magnitude plot for MIST isochrones in the K dwarf region.  Black points are the components of K+K wide binaries from our sample.  Different color lines represent different metallicity tracks. The three labels shown show the metallicity for the same colored lines.  Going from the +0.50 line to the -2.00 line, each different line represents a decrease of 0.25 in metallicity.  This increase to 0.5 after the -2.00 line.}
\end{figure}

To examine this doubling, \citet{2020hartman} focused on the K-dwarf regime and took a sample of 2227 K+K wide binaries with primary distances less than 250 pc and Bayesian probabilities $>99\%$ along with a cut based on color and absolute magnitude.  Additionally, they revised their definition of ``primary'' and ``secondary'' to be defined by $G_{BP} - G_{RP}$ color rather than \textit{Gaia} $G$ magnitude, with the component with a bluer color assigned as the primary star.  In this paper, we expand upon the previous analysis in \citet{2020hartman} by defining an expanded sample of K+K wide binaries from the 99,203 high probability wide binaries from the SUPERWIDE catalog.  In a similar manner to \citet{2020hartman}, we focus on stars withing a set color range $1.01 < G_{BP} - G_{RP} < 1.81$ as seen by the magenta boxes in Figure \ref{fig:priandsecselection}.  Like \citet{2020hartman}, our primary and secondary designations were initially determined by \textit{Gaia} $G$ magnitude, but are now changed to $G_{BP} - G_{RP}$ color; thus, we require each component to have both a $G_{BP}$ and $G_{RP}$ magnitude.  Unlike our previous work, we do not set a limit on absolute magnitude dependent on color.  We put limits so that the primary absolute magnitude is fainter than $M_G=4$ and the secondary magnitude is brighter than $M_G=9.4$; this eliminates red giants and white dwarfs from the subset.  In addition, we no longer restrict our sample based on distance, which greatly expands our selection.  We also lower our probability limit to 95\% as the majority of pairs in the 95\%$<P<$99\% are still real binaries.  The rationale for this shift is to include more binaries that are likely to be genuine wide binaries to increase the size of the sample for improved statistics.  This revised selection doubles the sample to 4947 candidate K+K wide pairs.  The color-magnitude diagrams for our selected binaries (with primaries in top panel and secondaries in bottom panel) are shown in Figure \ref{fig:priandsecselection}.  The magenta box represents our area of focus for this analysis and will be discussed below.  

\begin{figure}
\centering
\subfigure{
\includegraphics[width=3.3in]{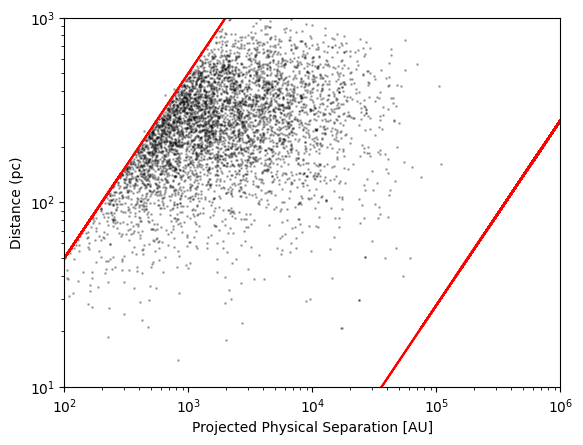}}
\subfigure{
\includegraphics[width=3.3in]{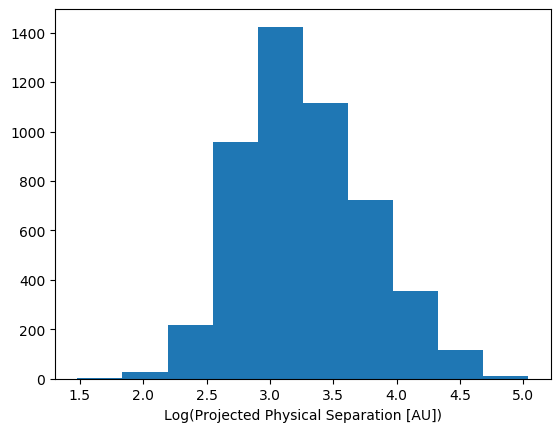}}
\caption{\label{fig:physsep}Top panel: Projected physical separation as a function of primary distance for the selected sample.  Most pairs reside between 100-600 pc.  Red lines indicate the 2\arcsec and 3600\arcsec cuts that are part of the SUPERWIDE catalog. Bottom panel: Distribution of projected physical separations for the selected sample.}
\end{figure}

A complicating factor in this analysis is the effect of metallicity difference between stars of similar mass, commonly known as ``cosmic scatter.''  To show how metallicity trends in the K-dwarf regime, we use the MIST isochrones \citep{2011paxton,2013paxton,2015paxton,2016dotter,2016choi,2018paxton} to model the metallicity tracks for solar age stars, with metallicity being $[Fe/H] = \log( (Fe/H)_{star} / (Fe/H)_{sun})$.  This is shown in Figure \ref{fig:overlumhr}.  We used the isochrones matching an age of $10^{9.65}$ years, $\frac{v}{v_{crit}} = 0.4$ and \textit{Gaia} magnitudes.  Starting from +0.5 in metallicity, these isochrones go to -4.0 in steps of 0.25 between +0.5 to -2.0 and then in steps of 0.5 between -2.0 and -4.0.  For most metallicity tracks down to -1.0, the metallicity tracks run parallel to each other in the K-dwarf regime.  In addition, we provide the distribution of projected physical separations as a function of distance and the histogram of projected physical separations for this sample in Figure \ref{fig:physsep}, showing that our binaries span a wide range of distances and physical separations.  The two red lines represent the effects of the angular separation cuts applied in \citet{2020hartman} at $2\arcsec$ and $3600\arcsec$, respectively.

\subsection{Light Curve Retrieval}
Starting with this sample of 4947 pairs, we cross-match our all primaries and secondaries with the TESS Input Catalog (TIC) by matching the \textit{Gaia} DR2 identification numbers in common between our sample and the TIC catalog using the modules astropy and astroquery \citep{2013astropy,2018astropy2,2019astroquery}.  Using the associated TIC ids, we use the Lightkurve package available in Python to search for TESS, K2 and Kepler targets and recover their light curves \citep{2018lightkurve}.  We retrieve all available sets of data for each target.  However, we examine each sector separately.  We first check if the target has available two-minute cadence light curves and download both the Simple Aperture Photometry (SAP) and Pre-search Data Conditioning SAP (PDCSAP) light curves.  If a target does not have two-minute data, we then check for data products from the MIT Quick Look-up Pipeline (QLP, \citealp{2020huanga,2020huangb}) and retrieve both the available light curves that resulted from their light curve extraction algorithm (SAP) and then light curves which underwent post-processing to remove stellar activity and instrumental noise which are referred to as KSPSAP .  If a star does not have either of those, we search for any K2 light curves and then finally for Kepler light curves.

We find 2928 primaries and 2494 secondaries that have light curves available from at least one of the missions.  Of those, 2463 are from the same binary system, i.e. we find 2463 pairs where a light curve is found for both the primary AND secondary component.  To be clear, each component has its own TIC number, however, in many of the cases, the light curves are essentially the same due to their proximity on the sky and the large size of the TESS camera pixels.  Our analysis finds that, of the 2463 pairs with both components having a light curve, 2047 have angular separations less than one TESS pixel (21\arcsec) while only 168 have angular separations larger than two TESS pixels (46\arcsec).  Therefore, the majority of the light curves for these systems blend the light from both stars unless a K2 or Kepler light curve is retrieved.  While this generally prevents us from knowing for sure from which of the components (primary or secondary) any signal is coming from, it still allows us to effectively search for eclipsing or rotating systems in both components at the same time.   Further ground-based photometric validation will however be needed to confirm which component is responsible for any modulation of the signal.  

In some cases, the signal may also be from a third star in close proximity to the pair.  However, we believe this to occur in a low number of our systems.  To examine this, we searched around each star in our sample for third stars within \textit{Gaia} DR2 that might contaminate the light curve.  We required that any other star near the searched star must not be its wide binary companion, have a \textit{Gaia} $G$ magnitude brighter than 18, and have a $\Delta G$ magnitude difference of less than 4.  We searched for stars within one TESS pixel (21\arcsec) and within two TESS pixels (42\arcsec).  We found that around 2\% had another star within one TESS pixel that could interfere with the signal from the target star and around 10\% had another star within two TESS pixels.  Due to the low probability of contamination indicated by these results, we will assume that if a periodic signal is detected in our analysis, it is from the target system and not a nearby field star.  This does not rule out the need for follow-up observations to confirm where the modulations are coming from for the reasons stated in the previous paragraphs.  

\subsection{Periodogram Analysis}
In order to identify systems with significant photometric modulations, we examine each of the light curves visually and sort them into one of four possible bins based on morphology: (1.) eclipsing/transiting system, (2.)fast rotators with periods less than five days, (3.) slow rotators with periods greater than five days and (4.)systems that show both rotation and eclipses; one example from each of these groups is seen in Figure \ref{fig:qlplcs}.  Eclipsing systems are easily identified from the dips in the light curves caused by another object passing in front of the target star while modulations from a fast/slow rotator show a sinusoidal/variable pattern which is due to spots on the surface of the star coming in and out of view.

We then conduct a periodogram analysis on these systems using the periodogram function from the Lightkurve package.  For the eclipsing systems and those systems that show both rotation and eclipses, we run this analysis twice using the Lomb-Scargle \citep{1982scargle} and Box-Least-Squares (BLS) \citep{2002kovacs} methods to identify and measure a period for the eclipses.  In most cases where the light curve is produced by the QLP, we use the KSPSAP fluxes.  This allows due to the removal of long term stellar trends for an easier calculation of the eclipsing binary's period.  Additionally, the rotation in these systems is found to be overwhelmingly in sync with the eclipses, for a good reason as tidal forces in close binaries will usually synchronize the rotation of both stars with their orbital period.  However, the light curves for several stars showed more noise in the KSPSAP flux than the SAP flux and, in these cases, we used the SAP flux to generate the light curves and the resulting phased light curves.  For the TESS two-minute, K2, and Kepler data, we use the PDCSAP fluxes to calculate the binary's period to take advantage of the cleaner data that the PDCSAP fluxes offer compared to the SAP fluxes.

After this first examination, the period is fine-tuned by hand to create a clean, phased light curve for each eclipsing/transiting system.  We conduct this analysis on the primary stars first and, if a secondary light curve for the same binary is found, we use the period of the primary to construct the phased light curves for the secondary if the same type of signal is seen.  We do note that no additional modeling has gone into this analysis as just identifying these systems as eclipsing/rotators suits the purpose of this study.

We run this analysis again on the systems that are identified as showing rotation.  In this case, we only use the Lomb-Scargle method as it can pick out the rotation signal better than the BLS method and we use only the SAP light curves unless the light curve is from K2 or Kepler, in which case we use the PDCSAP light curves.  This choice was made to avoid any potential cases where the detrending that is applied to the PDCSAP and KSPSAP from the QLP light curves could affect the stellar activity signals we hope to detect.  Additionally, we restrict the periods examined depending on whether the star was visually determined to have a rotation period greater than or less than five days from the visual analysis.  We again stress that just being identified as a fast or slow rotator suits the purpose of this study.  We leave it to a future paper to fully model the rotation signals in our identified rotators.  Much like the eclipsing/transiting analysis, if a secondary light curve is found with a primary counterpart, we use the primary's period to construct the phased diagram and see if this produces an acceptable phased light curve.  In any case, the identification of an accurate rotation period for a starspot signal is always elusive, as differential rotation in main sequence stars means that spots will move in and out of view at slightly different rates depending on their latitudes, and the period of the modulation will vary over time as the spot pattern changes.  Additionally, we find several systems where multiple variable signals are shown.  For these, we note both periods in our results but only count the system towards our analysis once.  

\section{Results}

In total, our search recovers 42 eclipsing/transiting systems, 16 systems showing both rotation and eclipses/transits, 105 systems showing rotation with a period less than 5 days, and 101 systems showing rotation with a period slower than 5 days.  We split these results into two groups.  In Tables \ref{tab:EBtruewidebinarycombined}, \ref{tab:Bothtruewidebinarycombined}, \ref{tab:FRtruewidebinarycombined}, and \ref{tab:SRtruewidebinarycombined}, we present the results for binaries which are two stars in the eyes of \textit{Gaia}.  In Tables \ref{tab:EBresolvedbinariescombined}, \ref{tab:Bothresolvedbinariescombined}, \ref{tab:FRresolvedbinariescombined}, and \ref{tab:SRresolvedbinariescombined}, we highlight the systems that are higher-order multiples in the eyes of \textit{Gaia} already.  These systems are discussed further in Section \ref{sec:resolvedhom}.  In each of these 8 tables, the columns are \textit{Gaia} DR2 ID Numbers, R.A. from \textit{Gaia} DR2, Decl. from \textit{Gaia} DR2, \textit{Gaia} $G$ magnitude, \textit{Gaia} $G$ flux error, NASA mission (either T-2min for TESS 2 minute light curve data, T-QLP for TESS data products from the QLP, K2 or Kepler), and mission ID (either TIC or Kepler catalog) for both components.  This is followed by the determined periods for components 1 and 2 with the final column being the SUPERWIDE angular separation.  Component 1 is the primary star as determined by bluer $G_{BP} - G_{RP}$ color and component 2 is the secondary for the true wide binary tables (\ref{tab:EBtruewidebinarycombined} - \ref{tab:SRtruewidebinarycombined}).  This assignment method is kept for the resolved triples; however, the other components are not taken into account.

\begin{figure*}
\centering
\subfigure{
\includegraphics[width=0.48\textwidth]{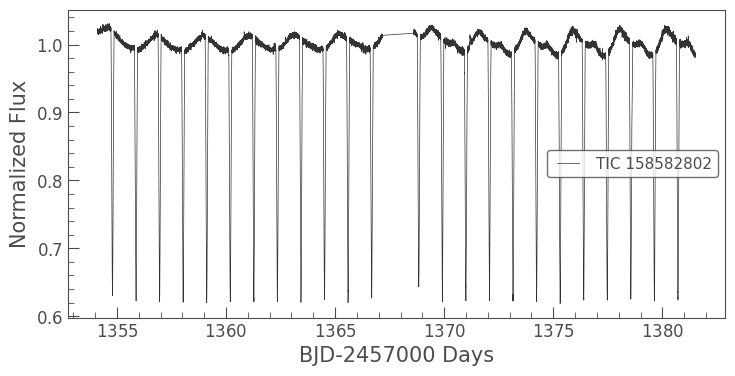}}
\subfigure{
\includegraphics[width=0.3\textwidth]{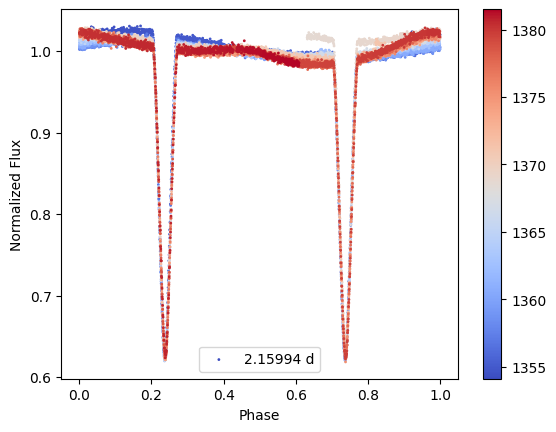}}
\subfigure{
\includegraphics[width=0.48\textwidth]{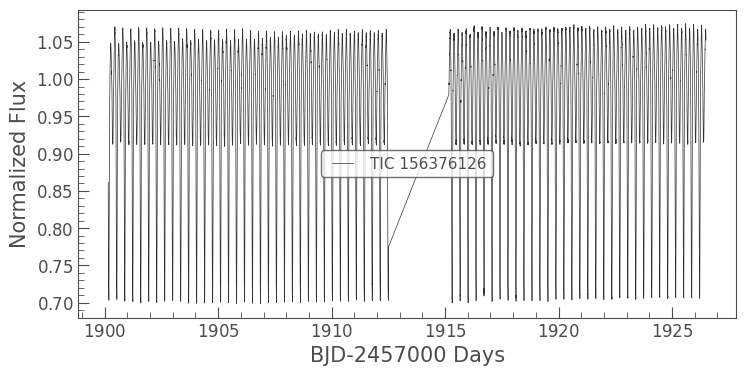}}
\subfigure{
\includegraphics[width=0.3\textwidth]{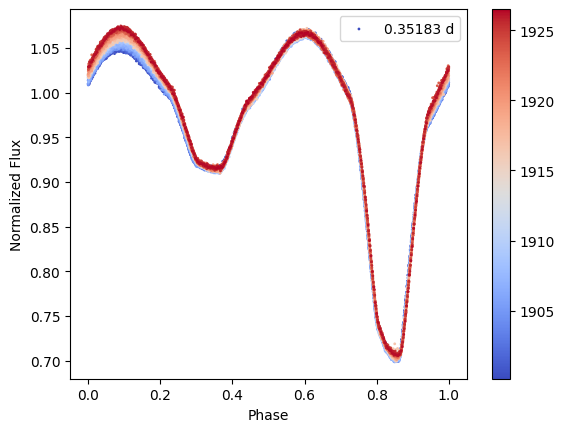}}
\subfigure{
\includegraphics[width=0.48\textwidth]{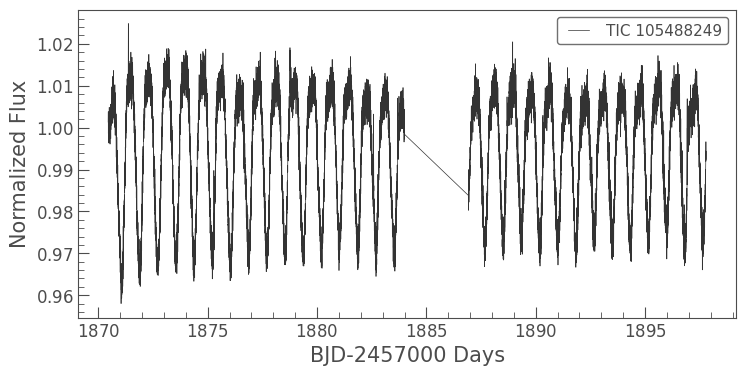}}
\subfigure{
\includegraphics[width=0.3\textwidth]{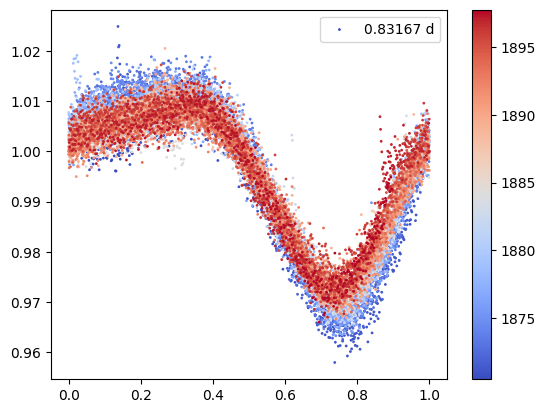}}
\subfigure{
\includegraphics[width=0.48\textwidth]{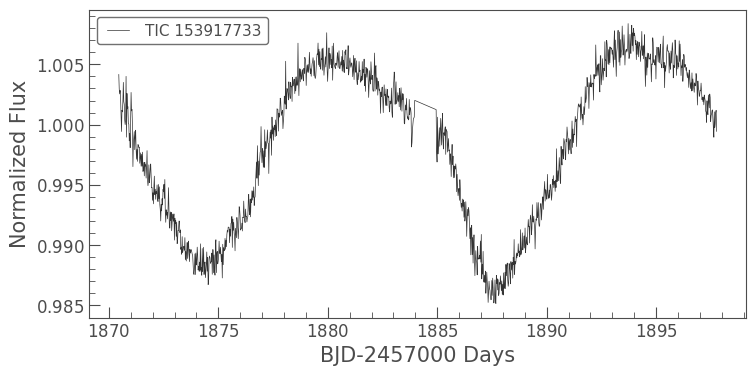}}
\subfigure{
\includegraphics[width=0.3\textwidth]{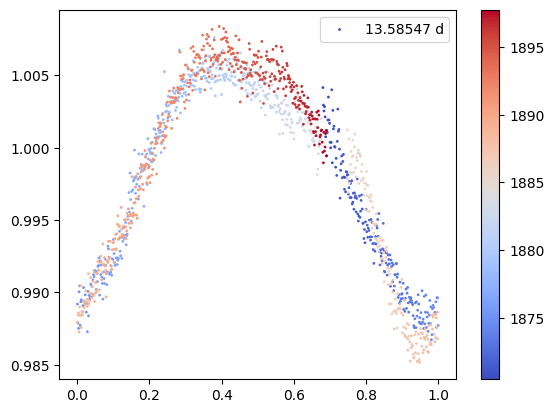}}

\caption{\label{fig:qlplcs}Examples of light curves and phased light curves for four primary stars in our sample.  Each plot pair shows the light curve obtained through Lightkurve on the left and the phased light curve on the right.  The period used for the phase folding is shown in the right panel for each star with the identifier for each star shown in the light curve plot.  Phased light curve points are colored by their time in the unphased light curve.  Top: Example of a system with both rotation and eclipses, Top-middle: Example of eclipsing system, Bottom-middle: Example of a system with rotation less than 5 days, and Bottom: Example of a system with rotation more than five days.  The different colored points and the color bar for the phased light curves show the epoch of the data points.}
\end{figure*}

In many of these cases, the light curves are the same for the two components except with slightly different base flux levels, which means that the TESS camera did not fully resolve the two components, and the measured light curve is that of the blended primary and secondary flux.  This means that although both components' light curves may show a variable signal, one of the components may not be variable.  Figure \ref{fig:qlplcs} shows examples of the original and phased light curves for stars from each of the four bins that we defined.  Within the tables, $P_1$, period of component 1 (primary), and $P_2$, period of component 2 (secondary), are determined in this manner.  If only one component has a variable signal detected in the light curve analysis, then that component is assigned the determined period from the light curve analysis.  If both components show variability and the period is different between the two, then both periods are reported.  Finally, if both components show variability and the period is the same, then the \textit{Gaia} flux error for each component is used to guess which component most likely houses the binary or stellar rotation signal with the component with the higher \textit{Gaia} flux error assumed to be the variable object.  Follow-up observations will be needed to confirm that our guess made here is correct. 

In a separate table, we highlight two wide binary systems where both components show signs of multiple sources of stellar modulation.  This is shown in Table \ref{tab:weirdfr}.  The two systems in question have evidence of either two or three stellar modulation signals in their periodograms.  We present the resulting periods in three columns of Table \ref{tab:weirdfr} with the rest of the columns being the same as in Tables \ref{tab:EBtruewidebinarycombined} - \ref{tab:SRresolvedbinariescombined}.  

\subsection{Testing the Lobster Diagram}
\begin{figure}
\centering
\includegraphics[width=3.3in]{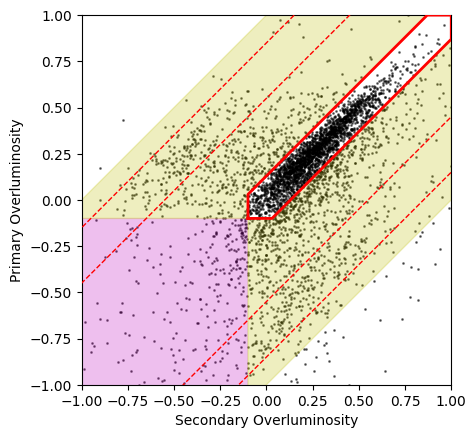}
\caption{\label{fig:lobtripsinc}The ``Lobster diagram'' plotting the overluminosity factor $F_{OL}$ of the primary star as a function of the $F_{OL}$ of the secondary star for the most of the 4647 K+K wide systems in the assembled sample.  Two key features stand out in this plot.  The dense locus along the 1:1 line represents ``true'' wide binaries where both components are single in the eyes of \textit{Gaia}.  Points lying outside of this locus represent systems where unresolved components make either the primary or secondary components suspiciously overluminous.  If a point falls in the purple region, it may represent a possible quadruple system or a young system, where both components are unusually overluminous.}
\end{figure}

As a test of the overluminosity correlation between primaries and secondaries in K+K wide binaries, we decided to re-examine the so-called ``Lobster diagram'' (Figure \ref{fig:lobtripsinc}) and assess where systems with eclipsing/rotator signals fall on the diagram.  \citet{2020hartman} defined the ``overluminosity factor'' ($F_{OL}$) as,
\begin{equation*}
F_{OL} = M_G - [M_G]_{Kref}
\end{equation*}
where $[ M_G ]_{Kref}$ is a reference value representing the absolute magnitude of a single-star K-dwarf of the same color with a set reference metal-abundance.  We adopt the same definition of the overluminosity factor,  ($F_{OL}$), with one key exception: in \cite{2020hartman}, the $M_G$ for each component was used separately in the calculation of $F_{OL}$.  However, we make a smaller but significant modification for this paper.  Since these are high probability wide binaries, their parallaxes should be nearly identical in most cases, as the orbital separations are much less than 1 pc.  It is thus fair to assume both stars are at the same distance, rather than introduce uncertainties from parallax measurement errors.  As such, for each binary, we adopt for both stars the parallax of the component that has the smallest parallax error in the \textit{Gaia} DR2 catalog.  We keep the same definition of $[ M_G ]_{Kref}$ as:
\begin{equation*}
[ M_G ]_{Kref} = 2.9 ( G_{BP} - G_{RP} ) + 2.5
\end{equation*}
This relationship is represented by the middle magenta line in both panels of Figure \ref{fig:priandsecselection}.  The line roughly tracks the division between the single star main sequence and the unresolved binary locus, although this choice is arbitrary.  A positive $F_{OL}$ means the component falls below the middle magenta line in Figure \ref{fig:priandsecselection}, while negative $F_{OL}$ means the component is above the line.  

The ``Lobster diagram'' exploits the fact that stars in wide binaries are expected to form within the same star-forming region and should fall on the same metallicity tracks on the H-R diagram.  In most cases, these tracks are expected to run parallel to the fiducial line we have defined.  In the ``Lobster diagram,'' the $F_{OL}$ values of the secondaries are plotted against the $F_{OL}$ of their associated primaries, with each pair represented by a single point.  Pairs with no unresolved companions should fall on a 1:1 locus because their $F_{OL}$ values will be similar.   On the other hand, the presence of an unresolved companion will increase the flux from one of the components (primary or secondary), and shift the point for that pair off the track, effectively revealing that the two components have values of $F_{OL}$ that do not agree with one another.  The magnitude of this shift will depend on the flux ratio between the star and its unresolved companion, which is related to the binary mass ratio, q, for main sequence stars.  

For this part of the analysis, we first remove any resolved higher-order multiples (to be discussed in a later section.)  This sets aside 300 pairs from our initial sample of 4947.  We show the overluminosity plot for most of the remaining 4647 wide binaries in Figure \ref{fig:lobtripsinc}, with 4268 systems falling within the shaded regions shown.  Several hundred pairs are not seen in this plot as one of the component's overluminosities falls outside of the plot limits.  The limits of the plot happen to correspond to the area bound by the magenta box in Figure \ref{fig:priandsecselection} and, for a point to appear in Figure \ref{fig:lobtripsinc}, both components must fall in the Figure \ref{fig:priandsecselection} magenta box.  This is examined more in Section \ref{sec:homf}.

There are four distinct regions in Figure \ref{fig:lobtripsinc}.  The area defined by the solid red lines, the body of the ``Lobster'', represents where pairs with two ``single'' components reside as their overluminosity factors are roughly the same.  This region is defined as the following:

\begin{equation*}
-0.1 < F_{OL,pri \& sec} < 1.0
\end{equation*}
\begin{equation*}
F_{OL,pri} > F_{OL,sec} - 0.13
\end{equation*}
\begin{equation*}
F_{OL,pri} < F_{OL,sec} + 0.13
\end{equation*}

This forms a ``true wide binary'' sequence composed of two single stars.  This was visually determined to match the observed density of points in this area with pairs with $F_{OL}>1.0$ and $F_{OL}<-1.0$ being considered suspicious (to be discussed later).  The two yellow shaded areas represent the areas where one component is unusually overluminous compared with its companion, as expected if it is an unresolved binary system, indicating the system is actually a triple.  Which way the pair deviates from the body of the ``Lobster'' determines which component is the possible unresolved binary.  If the primary is overluminous, then the pair will fall below the ``true binary'' (1:1) sequence whereas if the secondary is overluminous, then the pair will fall to the left of the sequence.  In both cases, the yellow areas extend away from the 1:1 line by 1 magnitude both below the line and to the left of the line for $F_{OL,pri/sec}>0.0$.  For $-0.1<F_{OL,pri/sec}<0$, only the area between $-1<F_{OL,pri/sec}<-0.1$ is considered as there was concern of sub-giant and giant contamination for pairs.  The purple shaded area represents the area where both components are unusually overluminous making the wide binary a possible quadruple system and is simply defined as a box with $-1<F_{OL,pri/sec}<-0.1$.  The regions denoted by the dashed red lines, the ``claws'' of the lobster, represent the area where unresolved systems with equal mass ($q\sim1$) are expected to cluster: these two regions are offset from the body by about 0.7 mag in overluminosity factor, which matches what is expected for an equal mass unresolved system on the color-magnitude diagram. In fact, the ``claws'' do show a definite over-density of objects, which is evidence for just such a sequence of unresolved, equal-mass close binaries.  Currently, the ``Lobster diagram'' only works well for wide binaries in the K-dwarf range, as the slope of the main sequence follows a linear color-magnitude relationship, sequences for stars of different metallicities have the same slope and the metallicity spread of the stars is relatively small, which is not the case for stars in the M-dwarf range, for example, where the color-magnitude-metallicity relationship is more complex.
 
\begin{figure*}
\centering
\subfigure{
\includegraphics[width=0.48\textwidth]{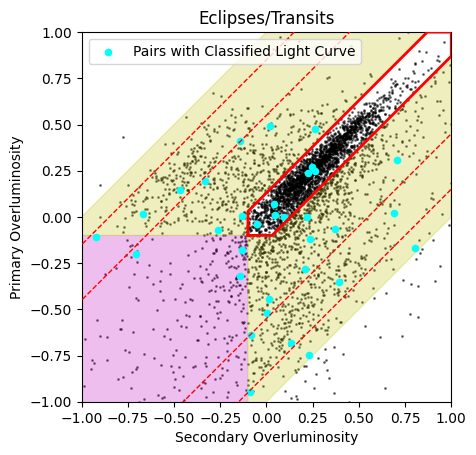}}
\subfigure{
\includegraphics[width=0.48\textwidth]{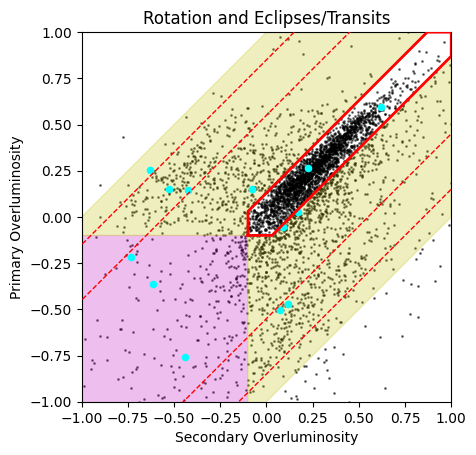}}
\subfigure{
\includegraphics[width=0.48\textwidth]{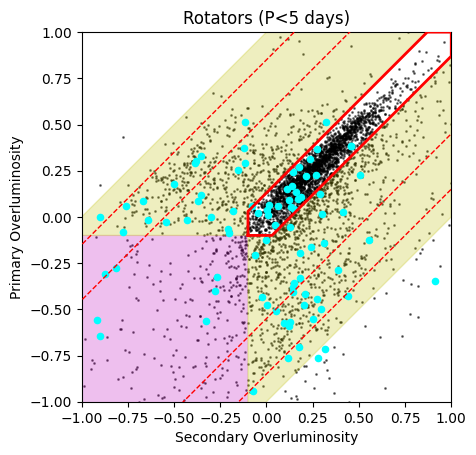}}
\subfigure{
\includegraphics[width=0.48\textwidth]{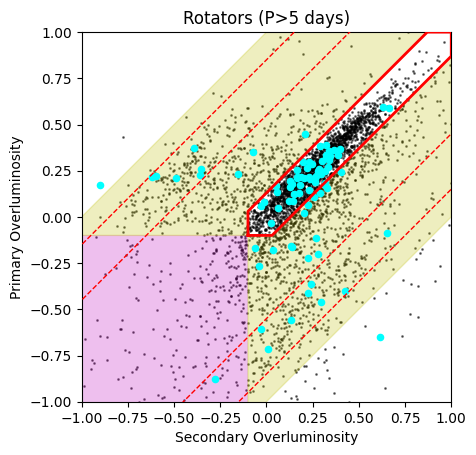}}
\caption{\label{fig:lobtess}Overluminosity plots highlighting the locations of TESS light curve targets (aqua points) that have been sorted into one of our four bins going from upper left to bottom right: (1.) systems where one component shows eclipses/transits, (2.) systems where one component shows both rotation and eclipses, (3.) systems where one component is a fast rotator ($P<5$ days), and (4.) systems where one component is a slow rotator ($P>5$ days).  Note that for the systems where a component shows either an eclipse or transit or is a fast rotator, most show at least one component to be overluminous, whereas slowly rotating systems are mainly in the ``true'' wide binary sequence.}
\end{figure*}

Using the binaries in Figure \ref{fig:lobtripsinc} to define the baseline, we plot out the locations of the eclipsing systems, systems showing rotation periods larger than 5 days (slow rotators), systems showing rotation periods less than 5 days (fast rotators), and systems showing both rotation and eclipses.  These are shown in the different panels in Figure \ref{fig:lobtess}.  22 pairs were not included on this plot, 5 eclipsing systems, 8 fast rotators, and 7 slow rotators.  For most of these systems, the reason for their exclusion was because one component's overluminosity was outside of the range of the plot and outside the magenta box in Figure \ref{fig:priandsecselection}.  As discussed further on, we only examine pairs which fall within the shaded regions of Figure \ref{fig:lobtess} for this analysis.  

Examining the four panels in Figure \ref{fig:lobtess}, we can verify if the ``Lobster diagram'' works as a possible method to determine if individual components of K+K wide binary systems are unresolved binary systems themselves.  In the plots of eclipsing systems and systems that show both rotation and eclipses  (top panels, Figure \ref{fig:lobtess}), we do confirm that the vast majority of systems with light curves classified in either of these two groups fall outside of the true wide binary sequence.  For the eclipsing systems, $78.9\%\pm20.7\%$ lie in the shaded regions of the top left panel of Figure \ref{fig:lobtess}, while $84.6\%\pm34.7\%$ lie in the shaded regions in the top right panel for the systems that show both rotation and eclipses with errors calculated using Poisson statistics.  We do not include systems that fall completely outside of the shaded regions or the true wide binary sequence.  For eclipsing systems that fall within the true wide binary sequence, there are two possible explanations.  First, the flux ratio between the companion that is eclipsing and the target star is small.  In this case, \textit{Gaia} would not pick up the excess flux from the faint component and the magnitude would be consistent with that of a single star.  Second, the light curve may be picking up eclipses from an unrelated background star caught in the TESS camera.  As mentioned previously, TESS pixels are large and can register the light from multiple stars in a single light curve; further follow-up would be needed to confirm this idea.  We also note that for those systems where only one component has a light curve that shows eclipses, the ``Lobster diagram'' correctly identifies whether the primary or secondary hosts the companion.  With these caveats, we believe that we have demonstrated that the ``Lobster diagram'' method is an efficient tool for identifying probable close companions in these wide binaries.  

Examining the light curves further, we examine the fraction of rotating stars that lie outside of the true wide binary sequence.  For the stars with rotation periods less than five days (lower left panel, Figure \ref{fig:lobtess}), $73.5\%\pm12.4\%$ lie in the shaded regions.  This significant result suggests a link between overluminosity and fast rotation, with the most likely explanation being that the fast rotation is caused by tidal synchronization of the rotation and orbital period in the close binary systems.  The fact that no eclipse is recorded simply indicates that the inclination of the system does not allow us to see the eclipses.  On the other hand, for stars with rotation periods greater than five days (lower right panel, Figure \ref{fig:lobtess}), only $42.0\%\pm8.6\%$ lie in the shaded regions, potentially pointing to their rotation and overluminosity not being related.  

\subsection{Additional Evidence from \textit{Gaia} and the Multiple Star Catalog}
\begin{figure}
\centering
\includegraphics[width=3.3in]{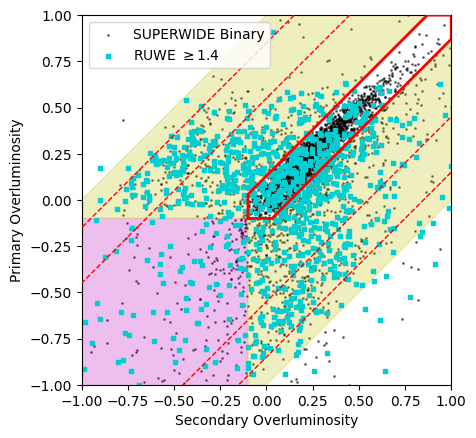}
\caption{\label{fig:RUWElob}``Lobster Diagram'' with pairs where one component's RUWE value from \textit{Gaia} DR2 is $\geq 1.4$ shown by the highlighted points.  The overwhelming majority of points lie outside of the true wide binary sequence (region bordered by solid red line).}
\end{figure}

To provide additional evidence that the ``Lobster diagram'' finds unresolved companions, we can examine the \textit{Gaia} data itself.  The Reduced Unit Weight Error (RUWE) parameter is an indicator of how good the astrometric fit is for a given star and can be used as a powerful tool to identify unresolved binaries within \textit{Gaia} \citep{2020belokurov,2021stassun,2022penoyre}.  Single stars with good astrometric fits typically have a RUWE between 1.0 and 1.4, although \citet{2021stassun} found evidence in a sample of eclipsing binaries systems that an eDR3 RUWE value in that range does not guarantee that a star does not have an unresolved companion.  As a test of the ``Lobster diagram,'' we recover the \textit{Gaia} DR2 RUWE values for our sample and, for each pair, we take the larger RUWE value from either component as the value for the system.  We then select the pairs with RUWE$\geq1.4$ as systems potentially hosting unresolved companions and plot them on the ``Lobster Diagram'' in Figure \ref{fig:RUWElob}.  Using the same method as in the previous section, $77.5\%\pm3.9\%$ of these systems reside outside of the true wide binary sequence agreeing with our findings from the light curve analysis.  

\begin{figure}
\centering
\includegraphics[width=3.3in]{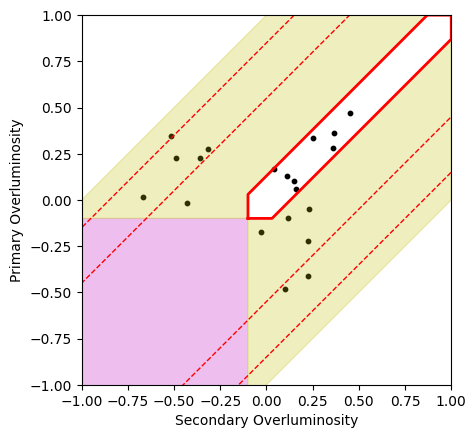}
\caption{\label{fig:msccrossmatch} Overluminosity plot for systems that are in SUPERWIDE as only two stars but are listed in the Multiple Star Catalog \citep{2018tokovinin} as having a known unresolved companion.  Systems outside the body have a bright unresolved companion that is detected as either a double-lined spectroscopic binary or a companion detected by speckle imaging.  Systems in the body are either faint companions detected by speckle imaging or single-lined spectroscopic binaries.}
\end{figure}

Finally, we cross-match our sample of 4647 systems with the Updated Multiple Star Catalog available on Vizier \citep{2018tokovinin}.  This catalog mostly consists of systems that are known to have 3 or more members.  We find 20 systems that are matches to the Multiple Star Catalog, all of them known to be triples with an unresolved third component.  Putting these 20 systems on the ``Lobster'' diagram in Figure \ref{fig:msccrossmatch} and applying the same criterion for the parallax used the calculation of the overluminosity for each component, we see that half of the systems fall on the true wide binary sequence, but the other half fall in the region where unresolved binaries are expected to reside.  Based on the data provided by the Multiple Star Catalog, the systems that fall in the unresolved binary regions mainly contain a component that is a double-lined spectroscopic binary while the ones that reside in the true wide binary sequence either contain a component that is a single-lined spectroscopic binary or a binary that is resolved with high angular resolution imaging.  As we are using the overluminosity of a component to tell if it has an unresolved companion, it follows that if the unresolved companion is faint, it will not contribute light to the system and will appear as a single star. 

\subsection{Resolved Higher-Order Multiples}
\label{sec:resolvedhom}
\begin{figure*}
\centering
\subfigure{
\includegraphics[width=0.48\textwidth]{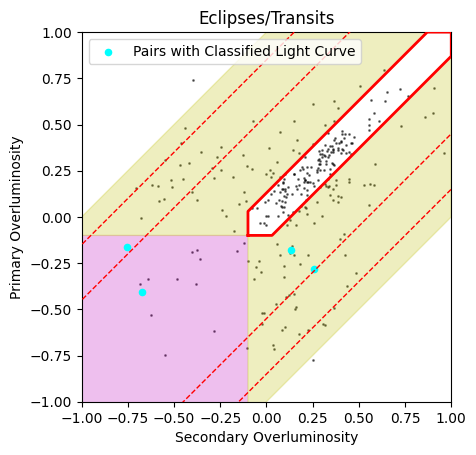}}
\subfigure{
\includegraphics[width=0.48\textwidth]{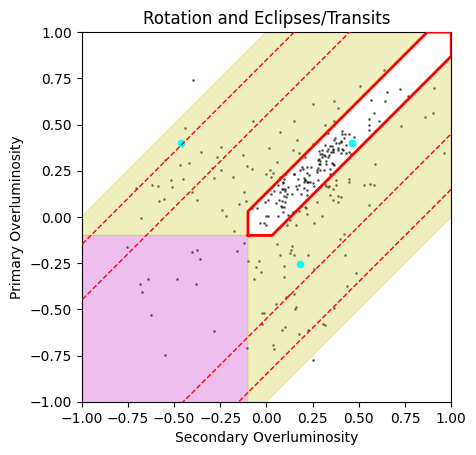}}
\subfigure{
\includegraphics[width=0.48\textwidth]{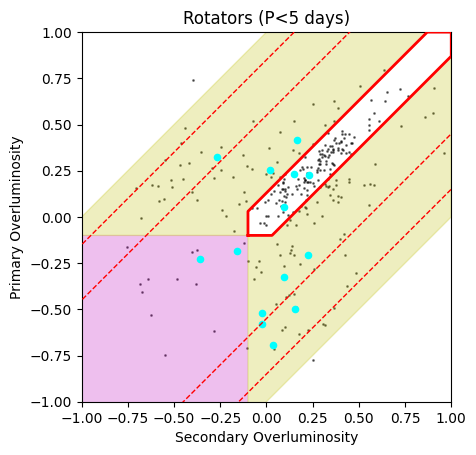}}
\subfigure{
\includegraphics[width=0.48\textwidth]{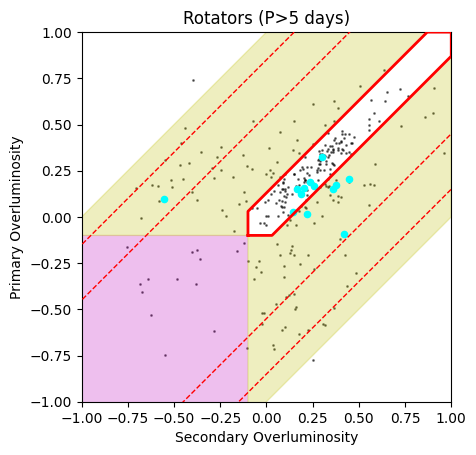}}
\caption{\label{fig:lobtriptess}Overluminosity plots highlighting the locations of TESS light curve targets for resolved higher-order multiples from \textit{Gaia} DR2.  Same format as in Figure \ref{fig:lobtess}. Some systems with TESS light curves fall outside of the range for this plot.}
\end{figure*}

In the section above, we deliberately excluded 300 ``pairs'' that we identified as being part of resolved higher-order multiple systems in \textit{Gaia} DR2; these are common proper motion systems of 3 or more resolved stars; each of these systems may contribute multiple "pairs" (since triples make 3 pairings, quadruples 6 pairings, etc.).  We now come back and address these systems as a group.  For our light curve analysis, we simply searched all 300 ``pairs'' for available TESS, K2 or Kepler light curves.  Among our detections, we find 4 systems with eclipsing binaries, 14 systems with a component showing rotation faster than five days, 13 systems with a component showing rotation slower than five days and 4 systems showing both rotation and eclipses/transits.  We plot the location of these pairs on the overluminosity plot in Figure \ref{fig:lobtriptess} using the same method as in Figure \ref{fig:lobtess}.  We only plot the K+K pairings of the resolved higher order multiples as that is what works on the lobster diagram, though some systems consist of all K-dwarfs (and contribute multiple ``pairs''), while others are K+K+something else (and contribute just one ``pair'' on the plot).  One of the systems showing both rotation and eclipses/transits falls outside of the window shown in Figure \ref{fig:lobtriptess}.  The same trends are observed showing that the majority of fast rotators and eclipsing systems fall outside of the true wide binary sequence while systems with slower rotation rates are more likely to be fall in the true wide binary sequence.

In Figure \ref{fig:lobtriptess}, we include all 300 ``pairs'' in the overluminosity plots.  However, for the next sections dealing with the higher-order multiplicity of K+K wide binaries, we only include a small number of higher-order multiples in our analysis.  The main reason for this is we want to only include resolved higher-order systems in those cases where if the system had been a wide binary with an unresolved companion, it would have made it into our sample.  This means we only include resolved systems consisting of a wide K+K pair and then a resolved closer K+something else system, where that something else is a K-dwarf or lower mass (M-dwarf).  Due to the way the SUPERWIDE catalog handles higher-order multiples, each possible pairing in a multiple system has its own entry in SUPERWIDE, i.e. a resolved triple will have 3 entries in SUPERWIDE, one for each pairing of components.  However, there are resolved higher-order systems in SUPERWIDE where there are only two entries; this is due to the $2\arcsec$ limit in SUPERWIDE so the two entries match to the ``pairs'' between the closer binary and the wide companion.  For these systems, we calculate the separation between the close binary in the system.  Taking these 300 ``pairs,'' we require each system to pass several cuts to be included in our analysis.  The projected physical separation of the K+K pair in the resolved system must not have the smallest separation of the ``pairs'' in the resolved triple, each ``pair'' in the resolved triple has a probability $>95\%$, and all other stars in the system must be K-dwarfs or lower mass.  These cuts result in a sample of 45 resolved higher-order multiples.

\subsection{Estimating the Higher-Order Multiplicity of K+K Wide Binaries}
\label{sec:homf}

One of the key features of the overluminosity plot is that it can be used to put a lower limit on the higher-order multiplicity of the K+K wide binaries based on the assumption that all pairs outside of the ``true'' wide binary sequence contain unresolved companions.  This will be a lower limit because not all stars with unresolved companions will appear as overluminous, as hinted from the light curve analysis which strongly suggests that unresolved systems with large mass ratios will not register as being overluminous.  To estimate the higher-order multiplicity fraction, we use the sample of 4647 binaries where the higher-order multiples have been excluded as they are already higher-order multiples. 

While in \cite{2020hartman} we counted everything that was not in the true wide binary sequence, this time we only count binaries as overluminous if they lie in the shaded regions of Figure \ref{fig:lobtripsinc}, i.e. we exclude objects that have such large overluminosity values in one component that they are considered suspicious, and assumed to be contaminants ($F_{OL,pri/sec} > 1.0$ and $F_{OL,pri/sec} < -1.0$).  There are two populations that are most impacted by this. Pairs containing sub-giant and giant stars are removed from the sample; They are assumed to be the points in Figure \ref{fig:lobtripsinc} that lie outside of the shaded regions in the lower right and pairs that are above the magenta box in Figure \ref{fig:priandsecselection}.  Metal-poor pairs with metallicities less than around -1.0 are also removed; they are assumed to be the points below the magenta box in Figure \ref{fig:priandsecselection}.  Both populations have large overluminosity values for two different reasons.  The sub-giants and giants appear overluminous as they are evolving off the main sequence with one component far above the main sequence and can be confused as unresolved binaries in our analysis.  The metal-poor stars appear overluminous because our assumption that the metallicity tracks are parallel to the straight line we defined as our reference breaks down for lower-mass, metal-poor systems.  This can be seen by comparing the -2.00 and +0.25 lines in Figure \ref{fig:overlumhr} across the entire K-dwarf region.  While they may be initially parallel from $1.0 < G_{BP} - G_{RP} < 1.1$, the slopes of the two lines rapidly diverge at larger $G_{BP} - G_{RP}$.  As our reference line for the calculation of $F_{OL}$ is roughly parallel to the metal-rich and solar metallicity tracks, this divergence breaks the correlation between the components of metal-poor stars and causes one or both of the components to have $F_{OL}>1$.  To avoid this issue, we remove 379 pairs which fall outside of the shaded regions in the ``Lobster Diagram.''

\begin{figure}
\centering
\includegraphics[width=3.3in]{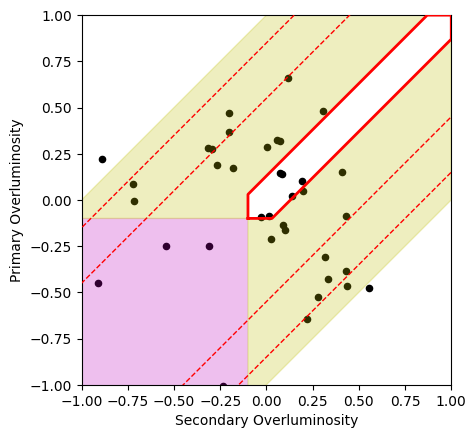}
\caption{\label{fig:resolvmultlob}Overluminosity plot for systems that are resolved higher-order multiples in SUPERWIDE, consist of a wide K+K pair, and do not include higher mass members.  For the calculation of the higher-order multiplicity, we only count systems which fall in the body of the ``Lobster Diagram'' or in the shaded regions.}
\end{figure}

With the remaining 4268 systems, we find that 259 fall in the purple shaded region representing possible quadruple systems, 1056 and 627 respectively fall in the yellow shaded regions below (primary component overluminous) and to the left (secondary component overluminous) of the true wide binary sequence, while the remaining 2326 systems fall in the true wide binary sequence.  Additionally, we include the resolved higher-order multiples that were removed from the analysis previously.  We take the sample of 45 higher-order multiples that were identified in Section \ref{sec:resolvedhom} and combine the pairs with the shortest separations.  This creates a sample of ``binaries'' where one or both components are treated as an unresolved system.  Most of the higher-order multiples are triple systems with 2 quadruple systems in a 2+2 configuration where both close binaries are K+M pairs.  We apply our analysis to these systems and show the resulting ``Lobster Diagram'' in Figure \ref{fig:resolvmultlob}.  We find that 3 systems are possible quadruple systems, 27 show signs of having an unresolved binary as one component and 7 fall on the true wide binary sequence.  Therefore, we count 30 of the higher-order multiples as overluminous and count 37 towards the total number.  8 systems fall outside of the shaded regions and are not included in the analysis.

\begin{figure}
\centering
\includegraphics[width=3.3in]{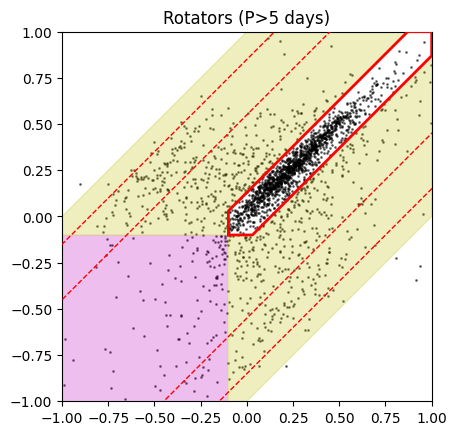}
\caption{\label{fig:lobnoevol}Overluminosity plot for K+K wide binaries with $G_{BP} - G_{RP} > 1.2$.  Note that the over-density of binaries directly below the true wide binary sequence that was present in Figure \ref{fig:lobtripsinc} is now absent.}
\end{figure}

Taking those systems of the 4268 that were not resolved higher-order multiples that fall outside of the true wide binary sequence plus the higher-order multiples discussed in the previous paragraph, we calculate a lower limit on the higher-order multiplicity fraction of $45.8\%\pm1.2\%$, with the error calculated with Poisson statistics.  This is roughly six percentage points higher than our $39.6\%\pm1.6\%$ estimate from \cite{2020hartman}.  We suspect this may be due to contamination from evolving (sub-giant) stars in the early K-dwarfs, $1.01 < G_{BP} - G_{RP} < 1.2$, which have been added after the removal of the distance limit used in \cite{2020hartman}, despite our best attempt at minimizing their presence (see above).  In Figure \ref{fig:priandsecselection}, there appears to be a steady increase in the density of points above the middle magenta line as one goes from $G_{BP} - G_{RP}$ of 1.2 to $G_{BP} - G_{RP}$ of 1.  This increase moves steadily further away from the magenta line as one goes to lower $G_{BP} - G_{RP}$ color.   This sub-population is also evident in Figure \ref{fig:lobtripsinc} as a concentration of stars located directly below the true wide binary sequence.  We believe that the majority of these binaries have a primary component that is beginning to evolve off the main sequence and is thus overluminous simply because of old age.  Figure \ref{fig:lobnoevol} shows the ``Lobster Diagram'' for the sample excluding binaries with $G_{BP} - G_{RP}<1.2$.  Using this smaller subset of 2251 binaries and including the 19 resolved higher-order multiples that satisfy this revised color cut (16 of which fall in the shaded regions), we determine a higher-order multiplicity fraction of $40.0\%\pm1.6\%$, roughly equivalent to our previous result.  

\subsection{Higher-Order Multiplicity as a Function of Projected Physical Separation}

\begin{figure}
\centering
\includegraphics[width=3.3in]{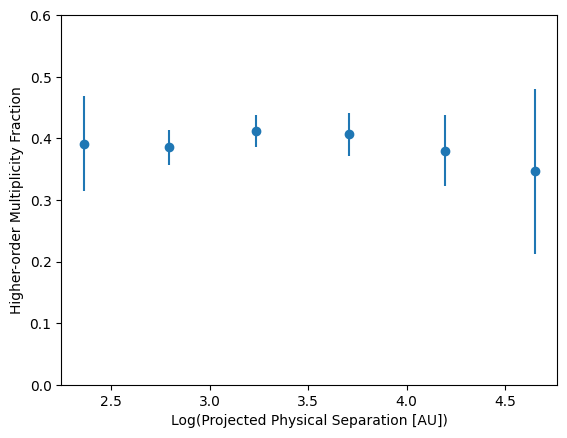}
\caption{\label{fig:physepanalysis}Higher-order multiplicity fraction as a function of projected physical separation for the widest component.  Error bars are determined by Poisson statistics.  No increase in higher order multiplicity is seen.}
\end{figure}

In \citet{2020hartman}, we compared the higher-order multiplicity fraction for the whole sample to a selection of extremely wide binaries with projected physical separations larger than 10,000 au.  We found that the fractions were similar to within the measurement errors.  For this paper, we expand this analysis to a wider range of physical separations.  Using the same criteria as was used to calculate the higher-order multiplicity fraction above and including the color cut to remove possible evolving star contamination, we split our sample into 6 bins with the first bin corresponding to all pairs with projected separations less than Log($\rho$)=2.5, going up in steps of 0.5 Log($\rho$) from Log($\rho$) = 2.5 to 4.5, and the final bin spanning all pairs with projected separations greater than Log($\rho$)=4.5.  The results are shown in Figure \ref{fig:physepanalysis} as a function of the average projected physical separation in each bin.  This reveals that the higher order multiplicity fraction is uniform at $\sim40\%$ and does not depend on the orbital separation of the widest component.  This is seemingly at odds with our current understanding of the higher-order multiplicity fraction as a function of projected physical separation, from which we would expect the higher-order multiplicity fraction to be increasing with separation, but this is not what we observe.  

\subsection{Higher-Order Multiplicity as a Function of Metallicity}
\begin{figure}
\centering
\includegraphics[width=3.3in]{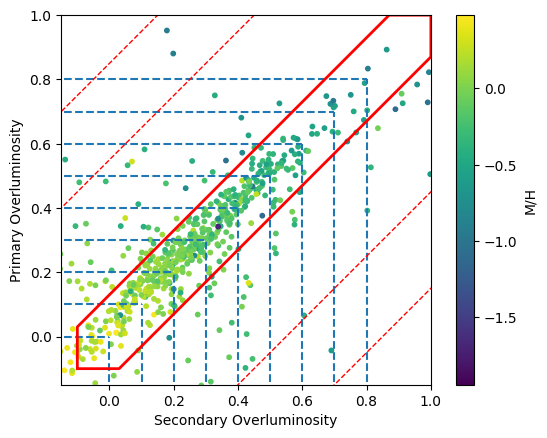}
\caption{\label{fig:lobmetal}Overluminosity plot for the primary and secondary stars which have photometric metallicities from \cite{2021medan} and have $G_{BP}-G_{RP} > 1.2$ .  Color scale represents the metallicity of the stars.  Dashed blue lines denote the metallicity bins used for the analysis.}
\end{figure}

One key feature that allows our study to pick out overluminous components in the K-dwarf region is that most of the metallicity tracks are parallel but offset from each other.  This enables one to identify possible overluminous components amongst the more metal-poor systems ($M/H\sim-1$), which might otherwise go unnoticed as overluminous metal-poor stars on their own are indistinguishable from more metal-rich stars.  Another consequence is that in the ``Lobster diagrams,'' metal-rich stars are found near the origin at (0,0) while metal-poor stars are shifted to higher $F_{OL}$ values (above and right).  To confirm this, we crossmatch our sample of 4647 K+K wide binaries with the catalog of \cite{2021medan} which provides photometric metallicities for a large sample of K-dwarfs, including most of the primary components in our catalog.  These metallicity values are shown color-coded in the ``Lobster diagram'' in Figure \ref{fig:lobmetal}. We observe a clear correlation between metallicity and overluminosity ($F_{OL}$) values. Using the photometric metallicity of the primary component as a guide, we split the overluminosity plot into eight equally spaced bins on the ``Lobster Diagram,'' starting from $F_{OL}=0.0$ and continuing to $F_{OL}=0.8$.  This is shown by the blue dashed lines in Figure \ref{fig:lobmetal}. 

\begin{figure}
\centering
\includegraphics[width=3.3in]{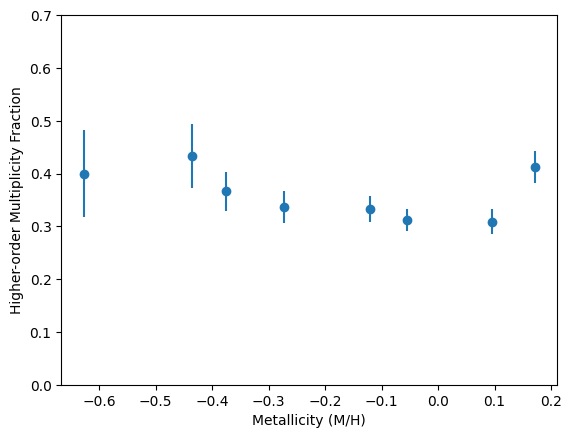}
\caption{\label{fig:metalfrac}Higher order multiplicity fraction as a function of metallicity.  Metallicity values for each point are determined by the average metallicity in the true wide binary sequence of the eight bins shown in Figure \ref{fig:lobmetal}.  Error bars are determined by Poisson statistics.}
\end{figure}

We redo our analysis and plot the resulting higher-order multiplicity fraction as a function of the average metallicity of the systems that fall in the true wide binary sequence for each bin in Figure \ref{fig:metalfrac}. If the primary does not have a photometric metallicity but the secondary does, then we adopt the secondary's value.  To reduce possible contamination from sub-giants as pointed out previously, we require both components to have $G_{BP}-G_{RP} > 1.2$.  We also include the 19 resolved triples which passed our cuts previously.  From Figure \ref{fig:metalfrac}, we see that the higher-order multiplicity fraction is consistent with having a uniform value over the metallicity range examined with the average value being $37.9\%\pm1.6\%$.  There is a hint of a downward trend with increased metallicity, or perhaps a modest increase in the highest metallicity bin, but it is not clear if these are significant.

\section{Discussion}

\subsection{Efficacy of the ``Lobster Diagram''}
As shown by Figure \ref{fig:lobtess} and Figure \ref{fig:RUWElob}, the overluminosity plot (the ``Lobster diagram'') enables one to identify many close companions that are unresolved in the \textit{Gaia} DR2 catalog, but can nonetheless be identified as close binaries.  The overwhelming majority of components with light curves that show eclipses/transits lie outside the true wide binary sequence, consistent with the expectation that unresolved binaries should be overluminous compared to main sequence stars of the same color.  For eclipsing systems that are found to lie along the true wide binary sequence, there are two possible explanations.  First, the companion that is eclipsing the target star could be faint, in which case the \textit{Gaia} photometry would not pick up the flux from the other star and the magnitude would be consistent with that of a single star.  Second, the light curve may be picking up the eclipses from an unrelated background star.  As mentioned previously, TESS pixels are large and can contain the light from multiple stars in a light curve.  Additionally, our RUWE analysis of the ``Lobster Diagram'' agrees with the light curve analysis in that the majority of systems with high RUWE values, assumed to be unresolved astrometric binaries, do lie outside of the true wide binary sequence.  Therefore, when this evidence is combined with the results of our light curve analysis, we conclude that appearing as overluminous on the ``Lobster diagram'' most likely indicates the presence of an unresolved companion in a wide binary system.

Of further interest is fact that the wide binaries with components that show rotation periods less than five days are also overwhelmingly found outside the true wide binary sequence.  This adds more evidence to the belief that fast rotation in older K-dwarfs is likely caused by tidal interactions with an unresolved close binary companion \citep{2020Angus}.  Stars spin down with age and high proper motion stars are expected to trend towards older age.  Therefore, these systems should be not rotating at periods less than five days, unless there is a close binary companion spinning them up.  As SUPERWIDE was created from a sample of high proper motion stars, there should be minimal contamination from young stars which would be the only way to get fact rotation in single stars.  As such, we believe these fast rotators are most likely to be binaries in which one component is spun up by the presence of an unresolved companion, which would explain both the fast rotation and overluminosity.  This echoes the findings of \citet{2019simonian} in their examination of fast rotators in the Kepler field which found that 59\% of stars with rotation periods less than seven days were overluminous. 

\subsection{Mass-ratio Limitation of the ``Lobster''}
While our analysis has shown that the ``Lobster Diagram'' can identify wide binaries with unresolved close companions, our method depends on \textit{Gaia} measuring an excess flux in the component with the unresolved companion.  This implies there should be a limiting mass ratio, q, below which an unresolved companion cannot be identified with our method.  To evaluate this limit, we revisit the MIST isochrones we examined in the beginning of the paper to model where unresolved companions of different q's and various metallicity values should fall on the overluminosity plot.  Figure \ref{fig:overlumhr} shows the distribution of MIST isochrones in the K-dwarf region with our sample of K+K wide binaries plotted in the background.  

\begin{figure}
\centering
\includegraphics[width=3.3in]{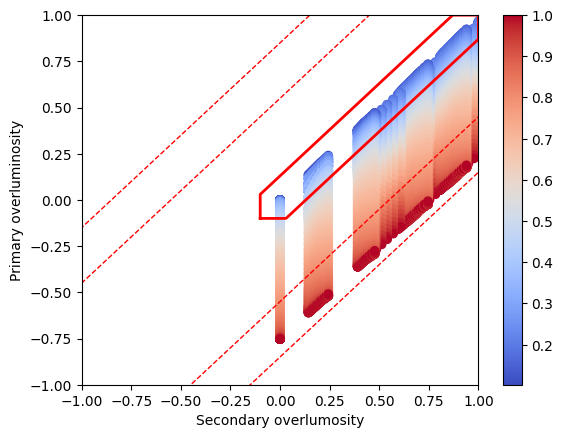}
\caption{\label{fig:MISTlob}Overluminosity plot using MIST isochrones to examine how mass ratio, q, affects the overluminosity for solar-age isochrones of different metallicities.  Color scale represents different q values.  We note that unresolved systems with a mass ratio above 0.5 are flagged as overluminous while mass ratios below this value fall within the true wide binary sequence.}
\end{figure}

We calculate the overluminosity factors for the primary and secondary components in this manner.  For a given solar-age isochrone with a fixed metallicity (a single line in Figure \ref{fig:overlumhr}), we step through the isochrone over the range of the K-dwarf region we used for this paper, $1.01\leq G_{BP}-G_{RP}\leq1.81$.  At each step, we obtain the mass and \textit{Gaia} $G$,$BP$ and $RP$ magnitudes provided by the isochrone.  We use these to calculate $F_{OL,pri}$ as,
\begin{equation*}
F_{OL,pri} = M_{G,double} - [M_G]_{Kref}
\end{equation*}
$M_{G,double}$ is the magnitude at the step in the isochrone plus additional light provided by an unresolved companion.  The value of the additional light is determined by iterating over all stars in the isochrone with a lower mass, meaning that for each step, we create how $F_{OL,pri}$ varies as a function of mass ratio.  $[M_G]_{Kref}$ is set as the value of the reference line at the color of the unresolved double.  For this part, we define our reference line as a cubic interpolation of the +0.5 metallicity isochrone, top line of Figure \ref{fig:overlumhr}.  For $F_{OL,sec}$, we set the overluminosity to be equal to that of the current step or to be equal to $F_{OL,pri}$ if it was a single star.  We do this for each metallicity isochrone seen in Figure \ref{fig:overlumhr} and plot the results in Figure \ref{fig:MISTlob}, with the color scale representing the value of q for each point.  Metallicity decreases monotonically along the 1-1 line in Figure \ref{fig:MISTlob} going from high metallicity on the left to low metallicity on the right.  As we defined our reference line by the +0.5 metallicity isochrone, the sequence originating at (0,0) on Figure \ref{fig:MISTlob} represents the q ratio distribution for one step on the +0.5 isochrone as $F_{OL,sec}=0.0$ for all steps on the +0.5 isochrone.  From Figure \ref{fig:MISTlob}, we conclude that we are sensitive to q ratios of roughly 0.5 or higher for most metallicities visible in this region of the overluminosity plot.

\subsection{Higher-Order Multiplicity Fraction and Projected Physical Separation}

Our results in Figure \ref{fig:physepanalysis} agree with our previous results from \citet{2020hartman} that the higher-order multiplicity fraction for the whole sample is comparable to that of binaries with projected physical separations larger than 10,000 au.  Figure \ref{fig:physepanalysis} indicates that the higher-order multiplicity fraction is uniform across the range of projected physical separation we are examining.  Past studies from \citet{2010raghavan}, \citet{2010law}, \citet{2014tokovininfgsample1}, \citet{2014tokovininfgsample2} and \citet{2019moemetal} have suggested that nearly half of the wide solar-type systems are part of triple systems and have suggested that this value increases with projected physical separation.  The difference in the overall multiplicity fraction (our 40.0\% vs. $\sim50\%$ from other studies) can be potentially explained by the fact the we know our study is missing some unresolved companions for several reasons.  First, the unresolved companions may be too faint for \textit{Gaia} to pick up and thus will not show up as overluminous.  Second, our method for constructing the SUPERWIDE catalog will bias our sample away from potential triple systems because if one of the components is an unresolved binary, there is a chance the unresolved companion will introduce significant errors in the \textit{Gaia} astrometric solution, resulting in a biased or inaccurate parallax and/or proper motion value.  If this causes the two wide components to have inconsistent parallaxes or proper motions, our search method will give these pairs a low probability.  Additionally, some pairs may be absent from our catalog because the unresolved companion may have caused the astrometric solution to fail, resulting in the star not having proper motion or parallax data from \textit{Gaia}.  A more in-depth study is needed to fully examine this possibility. 

Contributing to the difference in overall multiplicity fraction is the fact that our sample was constructed on the basis of both components of the wide binary falling within the K-dwarf region of the H-R diagram and represents one part of the larger wide binary population.  The previous studies of solar-type binaries mentioned above do not have this constraint and include binaries where the wide components have large mass ratios, i.e. a solar-type star with a wide M-dwarf companion.  These systems may have a different higher-order multiplicity fraction than systems where the wide components are similar masses.  We stress that our sample is a small part of the larger wide binary population.  Expanding our analysis presented here to lower masses to include K+M and M+M wide binaries is an ongoing project that may address this issue.

However, these explanations do not account for the lack of an increase in higher-order multiplicity as a function of projected physical separation.  Even with the issues mentioned above, one would expect the trend to still be seen unless it can be demonstrated that either unresolved companions cause a much more pronounced effect for larger separation binaries causing our search method to not identify them as possible pairs or the widest K+K systems have unseen companions with very low mass ratios that would be left undetected with our method.  This second option would also imply that moderately separated binaries tend to have third companions with higher mass ratios than their wider counterparts.  Even though our higher-order multiplicities are lower limits as explained above, we believe that the trend seen in Figure \ref{fig:physepanalysis} is real and indicates that the higher-order multiplicity of K+K wide binaries does not increase with physical separation.

\subsection{Higher-Order Multiplicity Fraction and Metallicity}
As shown in Figure \ref{fig:metalfrac}, when examined as a function of metallicity, the higher-order multiplicity fraction of K+K dwarf wide binaries is relatively constant.  There are some indications of variation over the metallicity range examined, manifesting as a slight decrease with increasing metallicity and a sharp increase at high metallicity, but we feel they are not statistically significant compared to the average value.  However, these indications do highlight the need for a closer look at this over a larger part of the wide binary population.  Examining this fraction over a wider range of metallicities would also be insightful as many of the recent studies examining either the close binary \citep{2019moemetal,2019elbadrymetal} or wide binary \citep{2021hwang} populations as a function of metallicity examine a larger range than we do here.  Even though these studies looked at the binary fraction rather than the higher-order multiplicity fraction primarily, there could be correlations between the two that can be explored if this analysis is expanded to higher and lower metallicities.

\section{Conclusions}
We have presented an analysis of 4947 wide pairs from the SUPERWIDE catalog looking for unresolved companions to wide binaries consisting of at least two K-dwarf stars with possible additional unresolved companions.  We search through TESS, K2 and Kepler light curves available through the MAST archive using Lightkurve, and supplement this with light curves produced by MIT's Quick Lookup Pipeline.  From this analysis, we recover 42 eclipsing/transiting systems, 16 systems showing both rotation and eclipses/transits, 105 systems showing rotation with a period less than 5 days, and 101 systems showing rotation with a period slower than 5 days.

Putting these systems on the ``Lobster diagram'', we find that the vast majority of systems which show eclipses, fast rotation, or both are overluminous.  We conclude that systems that appear overluminous on the ``Lobster diagram'' are most likely overluminous because of an unresolved companion.  Additional evidence is provided through an examination of where systems where one component has a high \textit{Gaia} DR2 RUWE value lies in the ``Lobster Diagram,'' with the majority falling in regions indicating they have unresolved companions.  Through a cross-match with the Updated Multiple Star Catalog \citep{2018tokovinin}, we find that double-lined spectroscopic binaries are also more likely to be found in areas of the ``Lobster diagram'' which indicate an overluminous component, while single-lined spectroscopic binaries are more likely to reside in the ``true'' wide binary sequence.  This points to the inherent drawback of examining overluminous systems for unresolved companions; that light from the companion needs to be seen to show overluminosity.  We investigate this using MIST isochrones finding that we should only be recovering unresolved companions if their mass ratios are greater than $\sim0.5$.  

Under our assumption that unresolved companions are responsible for the overluminous components in these wide binary components, we present a new estimate on the lower limit of the higher-order multiplicity of K+K wide binaries at $40.0\%\pm1.6\%$.  We finally examine the higher order multiplicity fraction of K+K wide binaries as a function of both projected physical separation and metallicity.  We find a uniform higher order multiplicity fraction with projected physical separation.  This is opposite of what many proposed formation scenarios suggest.  Additional trends with the metallicity of the pairs may also exist, but we do not see any statistically significant deviation from the average value over the range of metallicities examined.

\section{Acknowledgements}

We would like to thank the anonymous referee for all of their helpful and insightful comments.  ZDH would like to thank Gerard van Belle, Todd Henry, Douglas Gies, Xiaochun He, Catherine Clark, Bokyoung Kim, Maxwell Moe, Leonardo Paredes, Andrei Tokovinin, and Erika Wagoner for their insightful comments and help.  This paper includes data collected by the TESS mission. Funding for the TESS mission is provided by the NASA's Science Mission Directorate.  This work has made use of data from the European Space Agency (ESA) mission
\textit{Gaia} (\url{https://www.cosmos.esa.int/gaia}), processed by the \textit{Gaia}
Data Processing and Analysis Consortium (DPAC,
\url{https://www.cosmos.esa.int/web/gaia/dpac/consortium}). Funding for the DPAC
has been provided by national institutions, in particular the institutions
participating in the \textit{Gaia} Multilateral Agreement.  This paper includes data collected by the Kepler mission and obtained from the MAST data archive at the Space Telescope Science Institute (STScI). Funding for the Kepler mission is provided by the NASA Science Mission Directorate. STScI is operated by the Association of Universities for Research in Astronomy, Inc., under NASA contract NAS 5–26555.

Some of the data presented in this paper were obtained from the Mikulski Archive for Space Telescopes (MAST) at the Space Telescope Science Institute. The observations analyzed were obtained from publicly available TESS, K2, and Kepler data and the whole catalogs can be accessed via

\dataset[https://doi.org/10.17909/t9-st5g-3177]{https://doi.org/10.17909/t9-st5g-3177}

\dataset[https://doi.org/10.17909/t9-nmc8-f686]{https://doi.org/10.17909/t9-nmc8-f686}

\dataset[https://doi.org/10.17909/T9WS3R]{https://doi.org/10.17909/T9WS3R}

\dataset[https://doi.org/10.17909/T98304]{https://doi.org/10.17909/T98304} 

STScI is operated by the Association of Universities for Research in Astronomy, Inc., under NASA contract NAS5–26555. Support to MAST for these data is provided by the NASA Office of Space Science via grant NAG5–7584 and by other grants and contracts.

This work made use of Python and various python packages and services including, Jupyter, NumPy \citep{numpy}, SciPy \citep{2020SciPy-NMeth}, Matplotlib \citep{huntermatplotlib}, Astropy \citep{2013astropy,2018astropy2} and Astroquery \citep{2019astroquery}.  This research has made use of NASA's Astrophysics Data System.


\bibliography{sample631}{}
\bibliographystyle{aasjournal}



\begin{deluxetable}{lcccccclccccccccc}
\tabletypesize{\tiny}
\tablewidth{0pt}
\rotate
\setlength{\tabcolsep}{2pt}
\tablehead{
\colhead{\textit{Gaia} DR2 ID$_{1}$} & \colhead{R.A.$_{1}$} & \colhead{Decl.$_{1}$} & \colhead{$G_{1}$} &\colhead{$\sigma_{G Flux,1}$} & \colhead{Mission} & \colhead{Mission ID} & \colhead{\textit{Gaia} DR2 ID$_{2}$} & \colhead{R.A.$_{2}$} & \colhead{Decl.$_{2}$} & \colhead{$G_{2}$} &\colhead{$\sigma_{G Flux,2}$} & \colhead{Mission} & \colhead{Mission ID} & \colhead{$P_{1}$} & \colhead{$P_{2}$} & \colhead{$\rho_{wide}$}\\  
\colhead{} & \colhead{Degrees} & \colhead{Degrees} & \colhead{mag} & \colhead{electrons/s} & \colhead{} & \colhead{} & \colhead{} & \colhead{Degrees} & \colhead{Degrees} & \colhead{mag} & \colhead{electrons/s} & \colhead{} & \colhead{} & \colhead{Days} & \colhead{Days} & \colhead{\arcsec}
} 
\tablecaption{\label{tab:EBtruewidebinarycombined}Data on systems with eclipsing binaries. We provide the \textit{Gaia} DR2 ID, R.A., Decl., \textit{Gaia} $G$ magnitude, error on the $G$ flux and the mission and mission ID if a signal is detected.  In the last three columns, we provide the estimated period of the binary and the angular separation of the SUPERWIDE binary.  1 and 2 correspond to if the component is a primary or secondary respectively.  In many cases, the same signal is seen in both the primary and secondary light curves.  If the determined period is the same, the component with the higher flux error in \textit{Gaia} is assigned the period as a rough guess for which one is actually the binary.  Further follow-up is needed to confirm this.}
\startdata
528454567105351680&0.68838&66.80498&13.4171&31.1539&T-QLP&378537811&528454567105351936&0.68471&66.80352&15.3192&8.2184&--&--&1.74323&--&7.40068\\
523846239000492928&11.46881&63.0857&11.201&6870.704&T-2min&421110675&523846238991809920&11.46833&63.08497&14.87&41.2673&T-2min&421110675&0.31887&--&2.74123\\
4921052482695107712&11.67572&-54.62401&11.3451&227.6702&T-2min&281728276&4921052482695108096&11.67597&-54.62516&11.641&6190.275&T-2min&281728275&--&0.27337&4.19223\\
426104122059618176&13.471&59.69801&13.0162&42.0085&T-QLP&445321729&426104122059617792&13.47615&59.69844&13.1145&28.6916&T-QLP&445321730&4.95462&--&9.47929\\
346648360946112768&30.66204&43.81786&12.9613&48.1567&T-QLP&292004729&346648360946112896&30.66174&43.81712&16.431&23.1775&--&--&1.28905&--&2.77234\\
7265504117142400&44.8074&6.70864&13.0495&59.9898&T-QLP&387609082&7277225083276800&44.78046&6.71144&15.3885&8.3303&--&--&1.80803&--&96.82469\\
4456148829271936&47.21858&6.58576&12.3842&72.3589&T-QLP&365356083&4456153124271744&47.21815&6.58776&14.5364&99.579&--&--&0.35015&--&7.3574\\
4727302350444297984&47.55677&-57.7116&12.4138&1241.208&T-QLP&207200400&4727302350444297856&47.5576&-57.71094&14.2564&45.0618&--&--&0.26614&--&2.86924\\
249852541260745600&55.57063&50.40669&13.331&18.7244&T-QLP&428392461&249852545559867648&55.57038&50.40783&13.9512&15.1288&--&--&3.35286&--&4.14537\\
4681346647053604480&56.68396&-58.26176&14.0095&12.0222&T-QLP&197884431&4681346647053604608&56.68627&-58.26117&14.1452&277.8351&T-QLP&197884430&--&0.2569&4.86713\\
4888776799898384128&62.00957&-28.98742&13.9572&291.7718&T-QLP&44793390&4888776799898384000&62.00857&-28.98637&15.5375&5.2232&--&--&0.25018&--&4.92244\\
4791143534605894784&67.45739&-44.44902&12.1823&312.0932&T-QLP&301055977&4791143534605894912&67.456&-44.4483&13.0025&44.9007&T-QLP&301055978&0.43862&--&4.41067\\
4812652039332956288&76.11546&-42.08043&13.473&18.3414&T-QLP&200320624&4812652043629752320&76.11601&-42.08099&14.2319&24.4711&--&--&20.90979&--&2.49986\\
968501660527702912&95.14278&46.65993&13.3892&25.92&T-QLP&333872732&968501660527703168&95.14269&46.65932&16.2316&11.4078&--&--&3.76403&--&2.22502\\
3324516690989424512&96.48069&6.918&10.4745&554.8211&T-QLP&206707626&3324516725349160576&96.48012&6.92711&10.0089&12703.6925&T-QLP&206707639&--&0.26502&32.8443\\
5289371182736268416&116.37834&-61.30624&13.2538&1100.694&T-QLP&262610490&5289371182736267648&116.37735&-61.30785&13.9286&11.4283&T-QLP&262610486&0.24394&--&6.04814\\
1082870897246742016&117.37787&58.64735&12.0963&1315.2435&T-QLP&53330792&1082870901543250560&117.37612&58.64&13.9982&11.1743&T-QLP&53330798&0.25807&--&26.69194\\
1092459119677746560&128.7021&64.90385&12.4953&21.9454&T-QLP&802520259&1092459119679696384&128.70341&64.90346&14.1884&397.474&--&--&0.22804&--&2.45371\\
698598219065059456&138.03244&29.67713&14.116&11.5848&T-QLP&801338141&698598253424797696&138.0333&29.6766&14.0275&97.7094&T-QLP&801338142&--&0.74444&3.28841\\
5426877064682025856&138.4069&-44.3185&11.8355&60.7196&T-QLP&74970450&5426877068979062528&138.40782&-44.31867&14.1065&24.9439&--&--&3.16419&--&2.4404\\
5245551899175494656&153.57759&-65.37452&12.7758&26.4591&T-QLP&376899381&5245552620730005248&153.60666&-65.37348&12.8538&22.5927&--&--&7.00688&--&43.77432\\
5448690658119742592&156.48349&-32.26109&14.5277&6.7829&T-QLP&71715717&5448690658119742720&156.48369&-32.2617&16.2615&4.7758&--&--&0.95407&--&2.26042\\
5240469337981738240&165.73556&-64.01813&13.9177&14.7581&T-QLP&466391206&5240469337981817216&165.73133&-64.0175&14.8368&5.6762&--&--&5.29538&--&7.04878\\
5236430419447179776&177.00622&-66.11415&8.3367&3661.019&T-2min&410654298&5236430419447180544&177.00298&-66.11228&9.4783&969.744&T-QLP&410654317&9999.0&--&8.20646\\
3738936912850698624&202.77428&12.26034&10.4867&349.4072&T-2min&95525589&3739687539990111872&202.77151&12.26142&11.1434&7806.2939&T-2min&95525590&--&0.21801&10.47503\\
3616690087633221120&204.43974&-11.19255&13.0523&37.9124&K2&212572439&3616690465590343424&204.44144&-11.19221&14.6831&10.681&K2&212572452&2.58146&--&6.1269\\
5847439437677758464&211.69232&-68.77411&12.0685&2213.1644&T-QLP&448593526&5847439437674291584&211.68808&-68.77472&11.494&99.8279&T-QLP&448593523&0.24896&--&5.94516\\
5894442602013725952&217.76208&-54.58411&13.5508&452.3672&T-QLP&291662495&5894442602013723008&217.76344&-54.58638&14.9857&16.9919&--&--&0.30475&--&8.65459\\
5894118727130409472&219.08467&-55.25828&10.3986&296.6329&T-2min&411712609&5894118722803918464&219.08435&-55.26056&10.5303&3326.8658&T-QLP&411712622&--&0.29029&8.23962\\
5986270205398355840&236.19844&-48.07276&10.2884&309.7725&T-2min&267239205&5986270201075255936&236.19909&-48.07398&10.3631&388.3702&T-2min&267239205&--&0.83086&4.64307\\
1355675949894467840&249.6906&39.38875&11.8389&2151.5094&T-2min&68901227&1355675919831062016&249.68604&39.38543&14.1399&9.0798&T-QLP&68901224&0.27547&--&17.42781\\
2244830490514284928&303.38187&65.16233&12.4764&76.7925&T-2min&236887394&2244877876891663872&303.40829&65.21161&13.0176&33.5152&--&--&1.41997&--&181.83872\\
1964768420605269120&318.95511&37.71596&12.5745&38.0593&T-QLP&1961245902&1964768424911739264&318.95539&37.71525&12.7531&32.6193&T-QLP&1961245920&0.8956&--&2.66589\\
2613331778202503936&330.699&-12.31143&11.0738&3933.6979&K2&206091799&2613331782498358144&330.69805&-12.31177&13.9268&17.6119&K2&206091799&0.30677&--&3.58005\\
1895510015669888256&334.7517&30.64685&12.7531&45.5025&T-QLP&431119636&1895509946950411904&334.75273&30.64655&12.9479&621.1086&T-QLP&431127546&--&0.32423&3.36282\\
2401900777422950016&337.77862&-19.68317&12.5411&43.5541&T-2min&69818351&2401900777422949888&337.77564&-19.68297&12.2827&2684.2152&T-2min&69818350&--&0.2334&10.14599\\
2330058485800522240&351.54254&-29.6964&13.0883&1186.1339&T-QLP&270422783&2330058451440784768&351.52977&-29.69197&14.1614&19.1934&T-QLP&270422784&0.23013&--&43.00398\\
2011655689525260032&358.12146&60.48273&12.0842&36.3602&T-QLP&378355108&2011655689525260288&358.12502&60.48279&13.8049&12.5395&T-QLP&378355104&0.51843&--&6.31502
\enddata
\tablecomments{The binary in the system with \textit{Gaia} DR2 5236430419447179776 has a period of 9999.0 due to the presence of one eclipse in a single light curve.  This star is part of a known quadruple system \citep{2019tokovininspec}.}
\end{deluxetable}

\begin{deluxetable}{lcccccclccccccccc}
\tabletypesize{\tiny}
\tablewidth{0pt}
\rotate
\setlength{\tabcolsep}{2pt}
\tablehead{
\colhead{\textit{Gaia} DR2 ID$_{1}$} & \colhead{R.A.$_{1}$} & \colhead{Decl.$_{1}$} & \colhead{$G_{1}$} &\colhead{$\sigma_{G Flux,1}$} & \colhead{Mission} & \colhead{Mission ID} & \colhead{\textit{Gaia} DR2 ID$_{2}$} & \colhead{R.A.$_{2}$} & \colhead{Decl.$_{2}$} & \colhead{$G_{2}$} &\colhead{$\sigma_{G Flux,2}$} & \colhead{Mission} & \colhead{Mission ID} & \colhead{$P_{1}$} & \colhead{$P_{2}$} & \colhead{$\rho_{wide}$}\\ 
\colhead{} & \colhead{Degrees} & \colhead{Degrees} & \colhead{mag} & \colhead{electrons/s} & \colhead{} & \colhead{} & \colhead{} & \colhead{Degrees} & \colhead{Degrees} & \colhead{mag} & \colhead{electrons/s} & \colhead{} & \colhead{} & \colhead{Days} & \colhead{Days} & \colhead{\arcsec}
} 
\tablecaption{\label{tab:Bothtruewidebinarycombined}Data on systems which show both stellar rotation and eclipsing binaries.  Same format as Table \ref{tab:EBtruewidebinarycombined}.}
\startdata
2444056091886936320&3.67794&-5.21473&12.5811&266.5306&T-QLP&138691318&2444056091886936448&3.67872&-5.21533&14.1813&20.8062&T-QLP&138691316&0.87121&--&3.5447\\
2789335311745728128&12.75692&20.14043&12.3989&55.7862&T-QLP&435873807&2789335316039877504&12.75714&20.13939&11.8607&713.0345&T-QLP&435873808&--&0.7967&3.81103\\
397930923295485056&19.27049&45.90153&12.2928&54.0869&T-QLP&196844771&397925047780224256&19.27005&45.89917&12.2795&578.1299&T-QLP&196844769&--&1.31601&8.5929\\
4929615754128577920&21.85799&-49.47357&10.6629&1786.4357&T-2min&158582802&4929616132088007936&21.86097&-49.4708&11.0477&290.2174&T-2min&158582801&2.15994&--&12.17117\\
2452847168387507584&26.45556&-15.24315&13.1239&168.6245&T-QLP&382340348&2452847168387507200&26.45594&-15.24016&14.0227&22.3274&T-QLP&382340349&0.48418&--&10.85318\\
5712489301793743360&118.31403&-20.67203&11.6913&93.8926&T-QLP&142197564&5712488919536763520&118.31334&-20.67428&12.7944&24.0676&T-QLP&142197572&1.44742&--&8.44558\\
922165178219571328&119.67878&40.88004&12.0577&118.4929&T-QLP&371295954&922165182515948160&119.68393&40.87719&14.4556&11.0026&--&--&2.5462&--&17.40862\\
5309300208971922432&146.05378&-53.21069&14.3965&8.5056&--&--&5309300208971922176&146.05398&-53.21152&13.9207&76.9675&T-QLP&363156122&--&1.34688&2.98559\\
3750841638777083136&159.57006&-13.86663&13.3962&24.3894&T-QLP&386611784&3750841638777083264&159.56902&-13.86652&14.2725&65.4068&T-QLP&386611783&--&1.89349&3.64425\\
5908350565090411264&226.18569&-43.38032&13.3748&15.8828&T-QLP&335314651&5908350565090410624&226.18513&-43.38119&15.0913&6.4906&--&--&6.84818&--&3.45847\\
6634203199208276992&272.01175&-61.57752&14.1364&39.7239&T-QLP&303614824&6634203164848539392&272.00913&-61.57652&14.3616&6.7338&--&--&3.68014&--&5.76292\\
4538152265614974720&282.95933&27.04705&11.9351&106.1403&T-QLP&358236202&4538152265614974848&282.96106&27.04827&12.4417&36.6791&T-QLP&358236198&5.76513&--&7.08007\\
1818260149770354688&310.4318&22.50211&12.3451&75.1494&T-QLP&1943724735&1818260149765888384&310.43195&22.50276&12.3665&132.6982&T-QLP&1943724731&--&5.16639&2.40161
\enddata
\end{deluxetable}

\begin{deluxetable}{lcccccclccccccccc}
\tabletypesize{\tiny}
\tablewidth{0pt}
\rotate
\setlength{\tabcolsep}{2pt}
\tablehead{
\colhead{\textit{Gaia} DR2 ID$_{1}$} & \colhead{R.A.$_{1}$} & \colhead{Decl.$_{1}$} & \colhead{$G_{1}$} &\colhead{$\sigma_{G Flux,1}$} & \colhead{Mission} & \colhead{Mission ID} & \colhead{\textit{Gaia} DR2 ID$_{2}$} & \colhead{R.A.$_{2}$} & \colhead{Decl.$_{2}$} & \colhead{$G_{2}$} &\colhead{$\sigma_{G Flux,2}$} & \colhead{Mission} & \colhead{Mission ID} & \colhead{$P_{1}$} & \colhead{$P_{2}$} & \colhead{$\rho_{wide}$}\\ 
\colhead{} & \colhead{Degrees} & \colhead{Degrees} & \colhead{mag} & \colhead{electrons/s} & \colhead{} & \colhead{} & \colhead{} & \colhead{Degrees} & \colhead{Degrees} & \colhead{mag} & \colhead{electrons/s} & \colhead{} & \colhead{} & \colhead{Days} & \colhead{Days} & \colhead{\arcsec}
} 
\tablecaption{\label{tab:FRtruewidebinarycombined}Data on systems where rotation with periods less than 5 days is seen.  Same format as Table \ref{tab:EBtruewidebinarycombined}. Full table available online or upon request.}
\startdata
386604132462143360&0.14174&45.65775&12.8466&208.402&T-QLP&177699873&386604132462142208&0.12619&45.66734&15.214&5.5051&--&--&0.91178&--&52.17817\\
4994886612644765568&1.07508&-43.07455&12.4478&32.4989&T-2min&160147686&4994886612644765440&1.07712&-43.07646&13.107&142.8848&T-2min&160147687&--&0.87782&8.70426\\
2315484970275561600&6.59937&-33.78331&13.2312&154.2328&T-QLP&251840990&2315484974570922880&6.59704&-33.78219&14.2006&11.0911&T-QLP&251840989&2.10984&--&8.06516\\
5002803641824840064&12.61883&-35.38491&12.3131&497.1598&T-QLP&66376190&5002803646120723456&12.61559&-35.384&13.5461&16.8514&T-QLP&66376192&0.3063&--&10.05774\\
423411761676980480&14.21369&54.89546&13.4712&111.9598&T-QLP&604523784&423411761672430464&14.21313&54.89604&14.0055&10.0243&T-QLP&604523783&3.76287&--&2.39882\\
5003120614708227456&15.11848&-34.27271&14.3333&5.5541&--&--&5003120820866657408&15.12057&-34.26587&14.2147&29.5895&T-QLP&66471046&--&0.92062&25.37549\\
4989685372890146176&17.6714&-36.82419&10.7835&868.5528&T-QLP&183591689&4989685785207006080&17.69445&-36.7935&12.5217&31.1036&--&--&2.71714&--&128.91973\\
2534136811807695488&20.21302&-0.4445&12.905&164.0797&T-QLP&248944557&2534136811807695616&20.21427&-0.44704&14.8204&10.5621&--&--&1.80349&--&10.17767\\
4742592880694288512&31.94793&-56.27812&12.307&39.0397&T-QLP&231065372&4742592811974811904&31.95819&-56.28034&13.646&81.066&T-QLP&231065371&--&0.42644&22.01522\\
2513845668314348544&35.7859&2.76035&11.0857&1517.1899&T-QLP&420033303&2513839788503957760&35.77459&2.7553&13.3599&38.1214&T-QLP&420033299&3.46239&--&44.55011\\
5132158161175714048&38.5665&-18.44405&12.0973&485.1524&T-2min&64054033&5132158165471247360&38.56733&-18.44398&14.7023&25.447&--&--&2.52111&--&2.86393\\
...&...&...&...&...&...&...&...&...&...&...&...&...&...&...&...&...
\enddata
\end{deluxetable}

\begin{deluxetable}{lcccccclccccccccc}
\tabletypesize{\tiny}
\tablewidth{0pt}
\rotate
\setlength{\tabcolsep}{2pt}
\tablehead{
\colhead{\textit{Gaia} DR2 ID$_{1}$} & \colhead{R.A.$_{1}$} & \colhead{Decl.$_{1}$} & \colhead{$G_{1}$} &\colhead{$\sigma_{G Flux,1}$} & \colhead{Mission} & \colhead{Mission ID} & \colhead{\textit{Gaia} DR2 ID$_{2}$} & \colhead{R.A.$_{2}$} & \colhead{Decl.$_{2}$} & \colhead{$G_{2}$} &\colhead{$\sigma_{G Flux,2}$} & \colhead{Mission} & \colhead{Mission ID} & \colhead{$P_{1}$} & \colhead{$P_{2}$} & \colhead{$\rho_{wide}$}\\ 
\colhead{} & \colhead{Degrees} & \colhead{Degrees} & \colhead{mag} & \colhead{electrons/s} & \colhead{} & \colhead{} & \colhead{} & \colhead{Degrees} & \colhead{Degrees} & \colhead{mag} & \colhead{electrons/s} & \colhead{} & \colhead{} & \colhead{Days} & \colhead{Days} & \colhead{\arcsec}
} 
\tablecaption{\label{tab:SRtruewidebinarycombined}Data on systems where rotation with periods less than 5 days is seen.  Same format as Table \ref{tab:EBtruewidebinarycombined}. Full table available online or upon request.}
\startdata
4919427610667433728&0.03784&-57.48684&13.0809&64.8175&T-QLP&201236622&4919427610667433600&0.03676&-57.48719&16.5728&7.4764&--&--&13.08103&--&2.45258\\
4998071756096236160&5.0871&-38.5498&12.2084&415.2225&T-QLP&120609234&4998071756096142592&5.08575&-38.54934&12.4204&287.7033&--&--&7.18034&--&4.12958\\
4997625457454466048&6.67786&-38.91556&13.2156&58.0762&T-QLP&115445163&4997625461749785216&6.67559&-38.914&13.5356&16.1329&T-QLP&115444026&8.52143&--&8.50318\\
536946404645056384&9.81202&73.01276&13.1184&48.0427&T-QLP&275056961&536946404645064448&9.8199&73.00578&12.0824&51.2371&--&--&13.4356&--&26.45878\\
2555905080453468800&11.26911&5.55279&9.8996&999.2786&K2&220443444&2555905080453468544&11.26813&5.55497&10.4725&803.9951&--&--&7.47883&--&8.59294\\
405028575795255808&16.61153&53.20159&10.5104&607.362&T-QLP&240863503&405028683170219392&16.61537&53.20818&10.9395&526.1002&T-QLP&240863512&10.26844&--&25.15068\\
4983401560858949248&19.3624&-43.70162&11.6932&150.7562&T-QLP&229095201&4983401560858949120&19.36153&-43.70205&13.5102&44.381&--&--&13.67557&--&2.73911\\
410492981082524160&20.37908&53.75913&11.1635&383.769&T-QLP&623532078&410492981083566592&20.38006&53.7594&12.9304&175.7144&--&--&8.65604&--&2.30698\\
4915956589897978368&24.49648&-52.23626&13.2229&28.1294&--&--&4915956589897978240&24.48884&-52.23912&12.9144&59.8635&T-QLP&354606680&--&13.60253&19.74213\\
299039056489259520&28.11825&28.18108&12.3551&195.3365&T-QLP&28223779&299039193928212864&28.15283&28.18743&12.6349&87.817&--&--&5.90162&--&112.06324\\
4700919878172876032&30.85421&-63.95253&11.2631&196.706&--&--&4700921355641624192&30.88847&-63.93902&12.2925&97.9503&T-QLP&381205263&--&10.20376&72.79828\\
...&...&...&...&...&...&...&...&...&...&...&...&...&...&...&...&...
\enddata
\end{deluxetable}

\begin{deluxetable}{lcccccclccccccccc}
\tabletypesize{\tiny}
\tablewidth{0pt}
\rotate
\setlength{\tabcolsep}{2pt}
\tablehead{
\colhead{\textit{Gaia} DR2 ID$_{1}$} & \colhead{R.A.$_{1}$} & \colhead{Decl.$_{1}$} & \colhead{$G_{1}$} &\colhead{$\sigma_{G Flux,1}$} & \colhead{Mission} & \colhead{Mission ID} & \colhead{\textit{Gaia} DR2 ID$_{2}$} & \colhead{R.A.$_{2}$} & \colhead{Decl.$_{2}$} & \colhead{$G_{2}$} &\colhead{$\sigma_{G Flux,2}$} & \colhead{Mission} & \colhead{Mission ID} & \colhead{$P_{1}$} & \colhead{$P_{2}$} & \colhead{$\rho_{wide}$}\\ 
\colhead{} & \colhead{Degrees} & \colhead{Degrees} & \colhead{mag} & \colhead{electrons/s} & \colhead{} & \colhead{} & \colhead{} & \colhead{Degrees} & \colhead{Degrees} & \colhead{mag} & \colhead{electrons/s} & \colhead{} & \colhead{} & \colhead{Days} & \colhead{Days} & \colhead{\arcsec}
} 
\tablecaption{\label{tab:EBresolvedbinariescombined}Data on systems that are resolved higher-order multiples in \textit{Gaia} and have light curves which show eclipses. Same format as Table \ref{tab:EBtruewidebinarycombined}.}
\startdata
502304641542107904&79.68312&73.51948&12.6162&1716.466&T-QLP&140990375&502304641542108160&79.6787&73.52016&13.6979&14.8814&T-QLP&140990376&0.27666&--&5.14457\\
4031268437308995712&176.95394&35.22642&10.743&11820.5965&T-2min&156376126&4031268437308995584&176.96064&35.22732&12.8954&111.2375&T-QLP&156376127&0.35183&--&20.05945\\
6169396512667865216&202.44223&-32.33399&9.7661&27915.1888&T-2min&405533713&6169396443948387200&202.45015&-32.34049&10.7893&298.0736&T-QLP&405533708&0.27997&--&33.6048\\
2129813469650632064&292.57607&49.52598&10.8306&2984.1336&T-2min&26656583&2129813675806853760&292.56568&49.53592&10.4485&7046.7402&T-QLP&1882955197&--&0.52762&43.22962
\enddata
\end{deluxetable}

\begin{deluxetable}{lcccccclccccccccc}
\tabletypesize{\tiny}
\tablewidth{0pt}
\rotate
\setlength{\tabcolsep}{2pt}
\tablehead{
\colhead{\textit{Gaia} DR2 ID$_{1}$} & \colhead{R.A.$_{1}$} & \colhead{Decl.$_{1}$} & \colhead{$G_{1}$} &\colhead{$\sigma_{G Flux,1}$} & \colhead{Mission} & \colhead{Mission ID} & \colhead{\textit{Gaia} DR2 ID$_{2}$} & \colhead{R.A.$_{2}$} & \colhead{Decl.$_{2}$} & \colhead{$G_{2}$} &\colhead{$\sigma_{G Flux,2}$} & \colhead{Mission} & \colhead{Mission ID} & \colhead{$P_{1}$} & \colhead{$P_{2}$} & \colhead{$\rho_{wide}$}\\  
\colhead{} & \colhead{Degrees} & \colhead{Degrees} & \colhead{mag} & \colhead{electrons/s} & \colhead{} & \colhead{} & \colhead{} & \colhead{Degrees} & \colhead{Degrees} & \colhead{mag} & \colhead{electrons/s} & \colhead{} & \colhead{} & \colhead{Days} & \colhead{Days} & \colhead{\arcsec}
} 
\tablecaption{\label{tab:Bothresolvedbinariescombined}Data on systems that are resolved higher-order multiples in \textit{Gaia} and have light curves which show eclipses and rotation.  Same format as Table \ref{tab:EBtruewidebinarycombined}.}
\startdata
4703968201145328128&9.89361&-67.05368&13.8341&140.9912&T-QLP&38936416&4703956420051162752&9.86776&-67.06657&14.5812&9.8982&--&--&1.12926&--&58.89905\\
1151488123697266304&174.52794&87.31314&13.194&26.1128&T-QLP&900807400&1151488123697266432&174.51013&87.31312&13.936&16.9566&T-QLP&900807401&0.49101&--&3.00614\\
1151488123697266304&174.52794&87.31314&13.194&26.1128&T-QLP&900807400&1151488123697266688&174.58277&87.3196&13.8306&150.3899&T-2min&154808394&--&0.49093&25.0432\\
5789192087638234368&190.94676&-76.69667&12.9088&78.3116&T-QLP&360816293&5789192087637678592&190.95026&-76.69611&12.8861&42.268&T-QLP&360816296&3.55166&--&3.5235
\enddata
\tablecomments{\textit{Gaia} DR2 1151488123697266304, 1151488123697266432, and 1151488123697266688 are a resolved triple in which the light curves of each component shows the same variation.  We include all components of this system in this table.}
\end{deluxetable}

\begin{deluxetable}{lcccccclccccccccc}
\tabletypesize{\tiny}
\tablewidth{0pt}
\rotate
\setlength{\tabcolsep}{2pt}
\tablehead{
\colhead{\textit{Gaia} DR2 ID$_{1}$} & \colhead{R.A.$_{1}$} & \colhead{Decl.$_{1}$} & \colhead{$G_{1}$} &\colhead{$\sigma_{G Flux,1}$} & \colhead{Mission} & \colhead{Mission ID} & \colhead{\textit{Gaia} DR2 ID$_{2}$} & \colhead{R.A.$_{2}$} & \colhead{Decl.$_{2}$} & \colhead{$G_{2}$} &\colhead{$\sigma_{G Flux,2}$} & \colhead{Mission} & \colhead{Mission ID} & \colhead{$P_{1}$} & \colhead{$P_{2}$} & \colhead{$\rho_{wide}$}\\  
\colhead{} & \colhead{Degrees} & \colhead{Degrees} & \colhead{mag} & \colhead{electrons/s} & \colhead{} & \colhead{} & \colhead{} & \colhead{Degrees} & \colhead{Degrees} & \colhead{mag} & \colhead{electrons/s} & \colhead{} & \colhead{} & \colhead{Days} & \colhead{Days} & \colhead{\arcsec}
} 
\tablecaption{\label{tab:FRresolvedbinariescombined}Data on systems that are resolved higher-order multiples in \textit{Gaia} and have light curves which show signs of rotation with a period less than 5 days.  Same format as Table \ref{tab:EBtruewidebinarycombined}.}
\startdata
250324888882396544&56.75908&49.71261&10.8822&1670.0851&T-QLP&643654678&250324957598215168&56.75925&49.71318&11.9153&167.4713&T-QLP&643654684&3.84761&--&2.11277\\
3224017823514705920&84.44384&2.52385&10.6712&2663.4087&T-2min&323347748&3224016964521247616&84.43902&2.51581&10.8195&2966.6852&T-2min&323347760&--&3.13075&33.72459\\
268560491090059392&86.80975&56.50983&10.2078&1439.1232&T-2min&702593851&268561281363011328&86.77069&56.51083&10.3384&1304.0637&--&--&4.80719&--&77.67422\\
3166948080497672832&110.17059&14.29962&11.5602&107.2093&T-QLP&387343344&3166948080502189696&110.1711&14.30194&11.8106&338.3885&T-2min&387343345&--&4.29669&8.54597\\
3091076456717455232&122.46684&3.01932&12.0087&170.6519&T-2min&452981442&3091076456717454976&122.46721&3.01985&12.631&492.8716&--&--&0.51287&--&2.34251\\
5411671299904797696&148.77234&-45.92017&11.8626&1039.7486&T-QLP&35657528&5411671304209898496&148.77616&-45.91641&13.5643&15.9415&T-QLP&35657513&2.22181&--&16.57313\\
5371594208448932608&177.75679&-47.17904&12.045&81.0786&T-QLP&61879916&5371594204153503488&177.75783&-47.17941&13.2247&34.5699&--&--&3.7219&--&2.88668\\
3623208611038322816&200.00778&-11.36147&12.9144&114.6716&--&--&3623208606742706176&200.0072&-11.36152&12.4632&382.1829&K2&212564410&--&2.03467&2.05137\\
1698140428576723840&216.63811&70.31717&11.5792&852.5692&T-QLP&1001538841&1698140428576723712&216.64003&70.3168&13.3464&64.7363&T-QLP&1001538840&0.96294&--&2.66553\\
6201900168033571840&222.0395&-36.78403&9.6131&6450.1146&T-QLP&160059092&6201900172332643584&222.04326&-36.78359&10.22&1150.0637&T-QLP&160059095&3.85987&--&10.94953\\
1420728971964823168&259.258&56.20549&9.9395&1627.1327&T-2min&198413432&1420732300563539584&259.26041&56.21015&11.169&971.9233&T-2min&198413430&2.01754&--&17.46277\\
4582271887659609984&267.1949&25.95403&12.4935&189.7216&T-QLP&308175222&4582271887659610112&267.19484&25.95526&13.6148&15.8551&T-QLP&308175221&1.03185&--&4.41006\\
2203442575046718720&326.28919&59.69319&12.7431&147.161&T-QLP&388325838&2203442575046718336&326.28525&59.69462&13.5658&27.324&T-QLP&388325841&3.0095&--&8.80277\\
2206035326544213376&331.65084&63.75513&9.8475&4000.398&T-QLP&334810404&2206035326544214400&331.65314&63.7547&10.1755&2718.7122&T-QLP&2021747704&2.0651&--&3.96597
\enddata
\end{deluxetable}

\begin{deluxetable}{lcccccclccccccccc}
\tabletypesize{\tiny}
\tablewidth{0pt}
\rotate
\setlength{\tabcolsep}{2pt}
\tablehead{
\colhead{\textit{Gaia} DR2 ID$_{1}$} & \colhead{R.A.$_{1}$} & \colhead{Decl.$_{1}$} & \colhead{$G_{1}$} &\colhead{$\sigma_{G Flux,1}$} & \colhead{Mission} & \colhead{Mission ID} & \colhead{\textit{Gaia} DR2 ID$_{2}$} & \colhead{R.A.$_{2}$} & \colhead{Decl.$_{2}$} & \colhead{$G_{2}$} &\colhead{$\sigma_{G Flux,2}$} & \colhead{Mission} & \colhead{Mission ID} & \colhead{$P_{1}$} & \colhead{$P_{2}$} & \colhead{$\rho_{wide}$}\\ 
\colhead{} & \colhead{Degrees} & \colhead{Degrees} & \colhead{mag} & \colhead{electrons/s} & \colhead{} & \colhead{} & \colhead{} & \colhead{Degrees} & \colhead{Degrees} & \colhead{mag} & \colhead{electrons/s} & \colhead{} & \colhead{} & \colhead{Days} & \colhead{Days} & \colhead{\arcsec}
} 
\tablecaption{\label{tab:SRresolvedbinariescombined}Data on systems that are resolved higher-order multiples in \textit{Gaia} and have light curves which show signs of rotation with a period larger than 5 days.  Same format as Table \ref{tab:EBtruewidebinarycombined}.}
\startdata
2365377410625322112&4.32571&-19.33882&9.9305&1541.6159&--&--&2365377719862967936&4.35892&-19.34487&10.2913&534.5588&T-2min&399588417&--&5.15757&114.87933\\
323843493352143488&23.09436&39.7173&11.6719&230.4907&T-QLP&189470083&323842973662406528&23.07084&39.69025&13.387&30.3009&--&--&10.26843&--&117.16776\\
5046703262768137216&47.02469&-36.55834&11.9057&134.3107&T-2min&165161508&5046702987890214400&47.03953&-36.57441&13.3949&64.5274&--&--&8.7893&--&72.03567\\
66799763794633856&56.52719&24.56723&12.4231&378.1301&K2&211100398&66799768088857856&56.52886&24.56261&12.7025&117.7378&--&--&6.71026&--&17.50114\\
2889916543807416832&90.95986&-33.1103&11.1784&516.4169&--&--&2889916543807416960&90.96018&-33.11091&11.3107&338.0957&T-QLP&143276193&--&7.98163&2.37475\\
5262697541744463488&112.0563&-73.66145&12.3286&88.3562&T-2min&271808983&5262697546041903360&112.05987&-73.66148&12.6515&37.2045&T-2min&271808981&17.68226&--&3.61718\\
814027041396518144&141.97459&40.32156&12.3668&48.9078&T-QLP&16805478&814027041396512640&141.97531&40.32358&11.9826&99.7365&T-QLP&16805476&--&5.79829&7.52073\\
5242541333244980224&147.77316&-70.60837&11.6635&188.9318&T-QLP&371497079&5242541333244980352&147.77167&-70.60765&13.8806&36.7823&--&--&18.02215&--&3.14444\\
1662678945197227136&203.85108&59.91557&10.3135&1560.9607&T-2min&459225771&1662676467001283584&204.0655&59.88738&12.0314&855.4347&T-2min&160035228&5.53908&--&400.17947\\
6204912559313566464&223.19778&-33.61467&9.0624&2036.316&T-2min&48313985&6204920049736661760&223.33792&-33.43682&11.0684&480.616&--&--&6.09308&--&766.03355\\
1393462797287405184&228.47748&42.98933&10.6885&406.1471&--&--&1393462801582853760&228.4779&42.98884&12.9578&245.0947&T-2min&1101876681&--&10.66709&2.06712\\
4567896563401314304&257.45097&21.74666&10.8446&516.0353&T-2min&1309213495&4567896563401314432&257.45167&21.74666&12.6949&75.9084&--&--&8.98941&--&2.33176\\
6587386200247355520&329.79129&-35.95611&11.0342&438.4802&T-2min&197692963&6587386917506484480&329.76016&-35.9474&12.1578&105.4785&--&--&5.31188&--&95.99674
\enddata
\end{deluxetable}

\begin{deluxetable}{lcccccclcccccccccc}
\tabletypesize{\tiny}
\tablewidth{0pt}
\rotate
\setlength{\tabcolsep}{2pt}
\tablehead{
\colhead{\textit{Gaia} DR2 ID$_{1}$} & \colhead{R.A.$_{1}$} & \colhead{Decl.$_{1}$} & \colhead{$G_{1}$} &\colhead{$\sigma_{G Flux,1}$} & \colhead{Mission} & \colhead{Mission ID} & \colhead{\textit{Gaia} DR2 ID$_{2}$} & \colhead{R.A.$_{2}$} & \colhead{Decl.$_{2}$} & \colhead{$G_{2}$} &\colhead{$\sigma_{G Flux,2}$} & \colhead{Mission} & \colhead{Mission ID} & \colhead{$P_{1}$} & \colhead{$P_{2}$} & \colhead{$P_{3}$} & \colhead{$\rho_{wide}$}\\ 
\colhead{} & \colhead{Degrees} & \colhead{Degrees} & \colhead{mag} & \colhead{electrons/s} & \colhead{} & \colhead{} & \colhead{} & \colhead{Degrees} & \colhead{Degrees} & \colhead{mag} & \colhead{electrons/s} & \colhead{} & \colhead{} & \colhead{Days} & \colhead{Days} & \colhead{Days} & \colhead{\arcsec}
} 
\tablecaption{\label{tab:weirdfr}Data on two fast rotator systems which showed more than one stellar modulation.  We provide the \textit{Gaia} DR2 ID, R.A., Decl., \textit{Gaia} $G$ magnitude, error on the $G$ flux and the mission and mission ID if a signal is detected for each component.  In the last three columns, we provide the estimated period of the binary and the angular separation of the SUPERWIDE binary.  1 and 2 on these columns correspond to if the component is a primary or secondary respectively.  In this case, $P_1$, $P_2$ and $P_3$ refer to the detected periods in the systems.}
\startdata
4888246251178417280&60.01687&-29.04115&9.9958&2867.1958&T-2min&44670257&4888246251178417024&60.01633&-29.03795&9.7322&3221.0544&T-2min&44670258&0.31195&4.48792&--&11.62868\\
1372984431876717952&237.20106&36.35484&13.4462&110.2082&T-QLP&29265513&1372984431876717568&237.20344&36.35271&13.9078&103.3908&T-QLP&29265514&0.22236&0.44545&0.6331&10.31988
\enddata
\end{deluxetable}

\end{document}